\documentclass[12pt,a4paper]{article}
\usepackage[T1]{fontenc}
\usepackage[utf8]{inputenc}
\usepackage{authblk}
\usepackage{bbding} 

\usepackage{amsmath}
\usepackage{graphicx,float,color}
\usepackage{hyperref}
\hypersetup{colorlinks = true, linkcolor = blue, urlcolor  = blue, citecolor = blue, anchorcolor = blue}
\usepackage{multirow}
\usepackage[numbers,sort&compress]{natbib}
\usepackage{braket}
\usepackage{amsfonts}
\usepackage{amssymb} 
\usepackage{epic}
\usepackage{eepic}
\usepackage{epsfig}
\usepackage{latexsym}
\usepackage{color}
\usepackage{url}  
\usepackage{caption}
\usepackage[caption=false]{subfig}
\usepackage{float}
\usepackage{bm}
\usepackage{geometry}
\usepackage{slashed}
\usepackage[detect-all]{siunitx}

\usepackage{authblk}
\usepackage{physics}
\usepackage{amsmath} 
\usepackage{amsfonts}
\usepackage{color}
\usepackage{url}
\usepackage{graphicx,float,color}
\usepackage{verbatim}

\usepackage{tablefootnote}

\def\Mp{m_{\mathrm{Pl}}}
\def\lp{\ell_{\mathrm{Pl}}}

\def\Mmin{M_{\mathrm{min}}}

\usepackage{lipsum}
\makeatletter
\def\ps@pprintTitle{%
 \let\@oddhead\@empty
 \let\@evenhead\@empty
 \def\@oddfoot{}%
 \let\@evenfoot\@oddfoot}
\makeatother
\usepackage{etoolbox}

\begin{document}
\title{\textbf{Probing light scalars with not quite black holes}}    

\author[1,2]{Ximeng Li\thanks{liximeng@ihep.ac.cn}}
\author[1]{Jing Ren\thanks{renjing@ihep.ac.cn}}

\affil[1]{\normalsize Institute of High Energy Physics, Chinese Academy of Sciences, Beijing 100049, China}
\affil[2]{\normalsize School of Physics Sciences, University of Chinese Academy of Sciences, Beijing 100039, China}

\maketitle

\begin{abstract}

The rapid progress in gravitational wave astronomy has provided an opportunity for investigating the presence of long-range scalar forces that exclusively manifest around astrophysical black holes. In this paper, we explore a new possibility in this context, particularly in connection to the hypothesis that astrophysical black holes might be horizonless ultracompact objects (UCOs). In the absence of horizons, UCOs could feature unique interiors with extreme environments. This could help generate non-trivial scalar profiles and significant scalar charges. For demonstration, we consider 2-2-holes in quadratic gravity as a concrete example of UCOs. These objects can be formed by ordinary gases and 
closely resemble black holes externally. 
However, they have distinct interiors characterized by high curvatures and substantial redshift. In particular, the gases inside could reach extremely high temperatures or densities, making them an ideal object for investigating the generation of scalar profiles by UCOs.
Within a minimal model of the scalar field, we find that this unique environment enables the generation of a substantial scalar charge for astrophysical 2-2-holes, which is challenging for other stellar objects. 
The predicted scalar charge-to-mass ratio of 2-2-holes remains  nearly constant across a wide range of masses, offering different predictions for gravitational wave observations compared to other mechanisms. 
\\
\\
\textit{} 
\end{abstract}

\newpage
{
  \hypersetup{linkcolor=black}
  \tableofcontents
}

\section{Introduction}

Light scalar fields are commonly predicted in theories that go beyond the Standard Model (SM) of particle physics, as well as in various extensions of General Relativity (GR). 
In some instances, these scalar fields could potentially mediate additional long-range forces, thereby violating the weak equivalence principle in GR. For decades, extensive efforts have been made to search for such long-range forces through laboratory experiments~\cite{Chen:2014oda, Capolupo:2021dnl, Klimchitskaya:2017cnn, Chen:2022mzf} and astronomical observations of celestial objects~\cite{Williams:2004qba, Khoury:2003rn, Baessler:1999iv, Talmadge:1988qz, Tsai:2023zza}. These investigations have yielded null search results, thus imposing stringent constraints on the strength of potential new forces across a broad range of distances, spanning from the order of A.U. down to the micron scale (see Refs.~\cite{Adelberger:1991lgz, Fischbach:1992fa} for reviews).

The recent advancements in gravitational wave astronomy have provided new avenues for investigating the additional long-range forces. This is particularly important to search for scalar forces that manifest exclusively in the strong gravity regime. 
One generic mechanism underlying this phenomenon is the variation in the expectation value of the scalar field within compact objects compared to that in the weak gravity regime. As a result, a non-trivial scalar profile emerges in the strong gravity regime, leading to a nonzero scalar charge for an external observer situated at a far distance. Specific theoretical realizations of this concept have been actively studied in literature. One realization involves spontaneous scalarization in certain scalar-tensor theories of modified gravity~\cite{Damour:1996ke}. In this scenario, the non-minimal coupling of the scalar field to matter or spacetime curvature can lead to the formation of scalarized neutron stars~\cite{Damour:1996ke,Silva:2017uqg} or scalarized black holes~\cite{Silva:2017uqg, Cardoso:2013fwa, Doneva:2017bvd, Herdeiro:2018wub, Cunha:2019dwb,Herdeiro:2020wei}, depending on specific details (see Ref.~\cite{Doneva:2022ewd} for a review and further references on spontaneous scalarization).  
The second case considers the influence of finite density effects on QCD axions, which can give rise to scalarized neutron stars~\cite{Hook:2017psm, Sagunski:2017nzb}.
Other mechanisms include violations of a specific set of energy conditions~\cite{Heusler:1994wa, Bechmann:1995sa, Nucamendi:1995ex, Chew:2022enh} or the allowance for a time-dependent scalar field~\cite{Jacobson:1999vr, Barranco:2011eyw, Herdeiro:2014goa}.
From an observational standpoint, the cases predicting scalarized neutron stars face stronger constraints due to precise electromagnetic observations of pulsar binaries and the recent gravitational wave observation of neutron star binary inspirals by the LIGO-Virgo-KAGRA (LVK) collaboration~\cite{Damour:1996ke, Barausse:2012da, Shibata:2013pra, Shao:2017gwu,  Zhao:2019suc, Zhang:2021mks, Takeda:2023wqn}. Conversely, the increasing number of detected binary black hole mergers by the LVK and further observations involving supermassive black holes would play a crucial role in exploring 
non-trivial scalar profiles that exclusively manifest around black holes.

In this paper, we propose a new possibility concerning the sourcing of light scalar fields by astronomical black holes, specifically if these black holes are horizonless ultracompact objects (UCOs). This hypothesis of UCOs has recently garnered increased attention due to the potential for direct mapping of the immediate vicinity around black hole horizons through gravitational wave observations and the potential significance of near-horizon corrections in addressing associated theoretical challenges (see Ref.~\cite{Cardoso:2019rvt} for a review on UCOs).
An intriguing candidate for UCOs is the 2-2-holes~\cite{Holdom:2002xy,Holdom:2016nek}, representing a new family of solutions in quadratic gravity that may serve as the end point of gravitational collapse in the theory.
These objects closely resemble black hole in their exterior, but possess a distinctive interior characterized by extremely high curvatures  and a significant redshift. 
Depending on the specific models, the matter source residing within the deep gravitational potential of 2-2-hole interiors can possess an exceptionally high temperature or density~\cite{Holdom:2019ouz,Ren:2019afg,Aydemir:2021dan,Holdom:2022zzo}. This unique environment provides a promising opportunity to generate non-trivial scalar profiles of astrophysical black holes for a minimal model of the scalar field that features standard kinetic terms, minimal gravitational coupling, and is consistent with the necessary energy condition. This enables us to probe physics at extremely high energy levels that are typically inaccessible.
Furthermore, it offers an alternative perspective on the potential violation of the no-scalar-hair theorems for astrophysical black holes~\cite{Bekenstein:1972ny,Bekenstein:1995un} (see Ref.~\cite{Herdeiro:2015waa} for a review on the no-scalar-hair theorems). From an observational standpoint, the distinctive scaling behavior of the novel 2-2-hole interior may yield different phenomenological implications compared to other mechanisms.

This paper is organized as follows. In Sec.~\ref{sec:scalar}, we delve into the influence of environmental effects on the scalar potential for a minimal model, specifically focusing on high temperature corrections and high density corrections. 
In Sec.~\ref{sec:stellar}, we investigate the non-trivial scalar profile for stellar objects in the test field limit using the minimal model. We start by examining ordinary stellar objects and then shift our focus to 2-2-holes, exploring the implications of their differing characteristics in this specific context.
In Sec.~\ref{sec:GWobs}, we discuss the potential observational implications for scalarized 2-2-holes through gravitational wave observations.
We conclude in Sec.~\ref{sec:summary}.
The Appendix~\ref{app:backreaction} discusses the backreaction of the scalar field and the connection to no-scalar-hair theorems.

\section{Scalar potential with environmental effects}
\label{sec:scalar}

In a dense and hot environment, the effective Lagrangian for a real scalar $\phi$ with standard kinetic terms and minimal gravitational coupling can be expressed as
\begin{equation} \label{e1: L}
\mathcal{L}_{\phi}= -\frac{1}{2}\partial ^{\mu}\phi \partial _{\mu}\phi -V(\phi), \quad
V(\phi)=V_0(\phi)+V_T(\phi)+V_\rho(\phi)\,,
\end{equation}
where $V_0(\phi)$ denotes the scalar potential in the vacuum. The environmental effects are captured by $V_T(\phi)$ and $V_\rho(\phi)$, which account for the finite temperature and density corrections, respectively. These corrections are expected to become significant within stellar objects, altering the expected value of the scalar field and resulting in a non-trivial scalar profile that can be observed from the exterior of the stellar objects.

For illustrative purposes, we adopt a minimal model of the scalar field $\phi$ in this paper, utilizing the commonly used double-well potential
\begin{eqnarray}\label{eq:V0}
V_0(\phi)=-\frac{1}{2}\mu ^2\phi ^2+\frac{\lambda}{4}\phi ^4
=-\frac{1}{4}m_\phi^2\phi^2+\frac{m_\phi^2}{8\phi_0^2}\phi ^4\,,
\end{eqnarray}
where $\phi_0=\sqrt{\mu^2/\lambda}>0$ and $m_\phi^2=2\mu^2$ denote the vacuum expectation value (VEV) and the scalar mass.\footnote{Here, the vacuum is chosen as the minimum at $\phi=\phi_0>0$. In case the other minimum $-\phi_0$ is chosen, the subsequent discussion for the $F$-term case would remain the same if we flip the signs of $F$, $g_{\phi f}$, and $\delta\phi$ accordingly.}  This form of potential has also been encompassed in the improved no-scalar-hair theorems from Bekenstein~\cite{Bekenstein:1995un}. Therefore, this minimal case is sufficient to demonstrate how the no-scalar-hair theorems are violated in the context of not quite black holes. 
Moreover, in addition to the self-interaction, we also consider Yukawa coupling of $\phi$ to a Dirac fermion $\psi_f$, with
\begin{equation} \label{eq:Lf}
\mathcal{L}_f = 
\bar{\psi}_f\left( i\partial \!\!\!/-m_{f,0} \right) \psi _f
-g_{\phi f}\phi \bar{\psi}_f \psi _f\,.
\end{equation}
The fermion mass in the vacuum is given by $m_f=m_{f,0}+g_{\phi f}\phi_0$, where $m_{f,0}$ denotes the bare mass.

As we will discuss in detail later, when finite temperature or density corrections are significant, they can be effectively approximated by either a linear correction term or a quadratic correction term. The full potential can then be generally parametrized as 
\begin{eqnarray}\label{eq:fullV2}
V\left( \phi \right)&\approx& V_0(\phi)+F\phi+\frac{1}{2}G\phi^2\nonumber\\
&\approx& F\phi -\frac{1}{4}\left( m_\phi ^2-2G \right) \phi ^2+\frac{m_\phi ^2}{8\phi _{0}^{2}}\phi ^4\,,
\end{eqnarray}
where $F$ and $G$ represent the coefficients for linear and quadratic terms, respectively. For convenience, they are referred to as the $F$-term and $G$-term cases, respectively. 
In the scalar field equation of motion (EOM), the linear term introduces an effective force and the quadratic term leads to corrections to the effective mass.

For an intuitive understanding, it is helpful to consider the scalar field EOM as an equation that governs the time evolution of a particle along a single spatial direction in classical mechanics, where $-V(\phi)$ becomes the effective potential for the particle. A non-trivial scalar profile can be expected if a new maximum of $-V(\phi)$ higher than the one at vacuum is developed in the stellar interior due to environmental effects. In particular, the resulting scalar profile corresponds to the particle starting to fall off  slightly away from the new maximum at the initial time and stopping right at the vacuum at infinity due to friction.

\begin{figure}[H]
  \begin{minipage}{0.5\textwidth}
  \centering
  \includegraphics[height=5.cm]{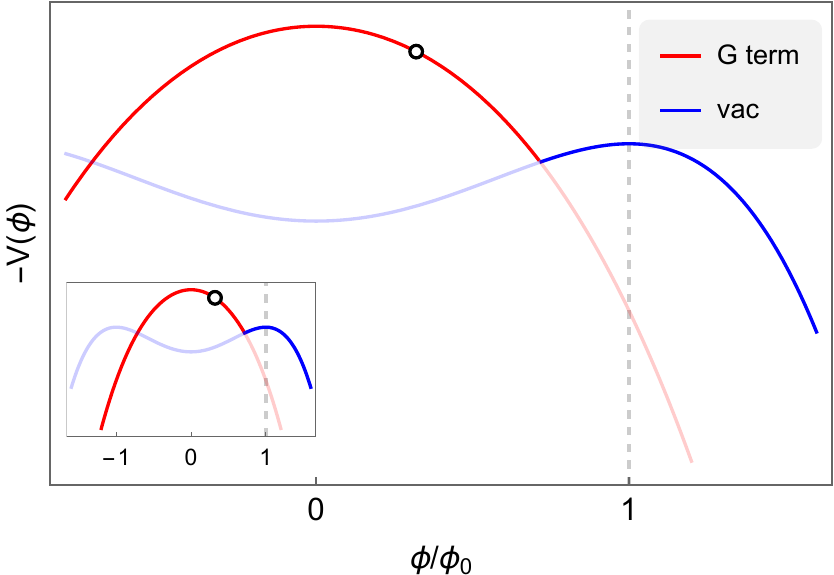}
  \end{minipage}
  \begin{minipage}{0.5\textwidth}
  \centering
  \includegraphics[height=5.cm]{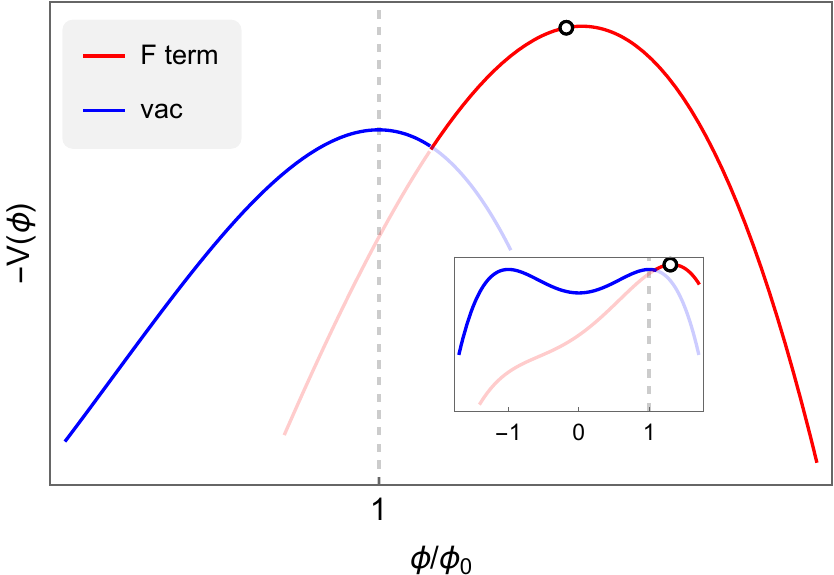}
  \end{minipage}
    \caption{Schematic illustration of two different ways for obtaining a non-trivial scalar profile with environmental effects. In both panels, the red and blue curves represent the potential with and without the environmental corrections. The open circle denotes the boundary value of $\phi$ at $r=0$, and the vertical dashed line represents the VEV that $\phi$ asymptotically approaches at spatial infinity. Left: the $G$-term case (quadratic correction)  with $G\gtrsim m_\phi^2/2$, where the scalar field moves to $\phi_0$ from the left. Right: the $F$-term case (linear correction) with $F<0$, where the scalar field moves to $\phi_0$ from the right.}
  \label{fig:VphiFG}
  \end{figure}

Figure~\ref{fig:VphiFG} provides a schematic illustration of the effective potential for the particle. In the $G$-term case, a new maximum of $-V(\phi)$ can be obtained at $\phi=0$ if the quadratic term flips the sign with $G\gtrsim m_\phi^2/2$. A non-trivial scalar profile would then develop, with $\phi$ evolving from some value slightly above zero to the VEV from the left. In the $F$-term case, a new maximum, higher than that at the VEV, develops at $\phi>\phi_0$ and smoothly moves away from $\phi_0$ for a negative $F$. This is in contrast to the other case, where $G$ has to exceed a certain threshold value to enable the existence of the new maximum. A non-trivial scalar profile develops for this case, with $\phi$ evolving from some value slightly below the new maximum to the VEV from the right.
It is worth noting that in both scenarios, the scalar field EOMs are non-linear, thereby precluding their analysis as eigenvalue problems. This contrasts with the analysis of scalarized black holes within the context of quadratic scalar-Gauss-Bonnet gravity in Ref.~\cite{Silva:2017uqg}, where the EOM is linear and a non-trivial scalar profile emerges only for discrete values of the associated coupling parameter.

Next, we will derive the explicit forms of $F$ and $G$ from the environmental effects.
For the finite-temperature effects, we consider one-loop corrections
\begin{eqnarray}\label{eq:VT}
V_T(\phi, T)=I_B(m_\phi(\phi),T)+4 I_F(m_f(\phi),T)\,,
\end{eqnarray}
where the contribution from the bosonic and fermionic degrees of freedom are given by~\cite{Dolan:1973qd}
\begin{eqnarray}
I_{B,F}(m_i(\phi),T)\equiv \pm\dfrac{T^4}{2\pi^2}\int_0^{\infty} \mathrm{d}y\;y^2\ln\left[1\mp\exp\left(-\sqrt{y^2+\dfrac{m^2(\phi)}{T^2}}\right)\right]\,,
\end{eqnarray}
where the $+$ ($-$) and $-$ ($+$) sign at the front (in the integrand) are for bosons (fermions). $m_i(\phi)$ denotes the field dependent mass, with $m_\phi^2(\phi)=-\frac{1}{2}m_{\phi}^2+\frac{3}{2}m_{\phi}^2\phi^2/\phi_0^2$ and $m_f(\phi)=m_{f,0}+g_{\phi f}\phi_0$.
Note that we ignore the contribution from the resummed thermal daisy loops for the bosonic Matsubara zero modes in Eq.~(\ref{eq:VT}), i.e. $J_B(m_\phi(\phi),\Pi_\phi, T)\equiv \frac{1}{12\pi}T [m^3_\phi(\phi)-(m^2_\phi(\phi)+\Pi_\phi)^{3/2}]$, where $\Pi_\phi\propto T^2$ is the thermal mass~\cite{Carrington:1991hz,Arnold:1992rz}. 
In cases where $V_T$ is non-negligible, i.e. in the high-temperature expansion, we have confirmed that the contribution of $J_B$ is significantly smaller than that of other terms.

Although the close form of $V_T(\phi, T)$ in Eq.~(\ref{eq:VT}) is absent, there are good approximations at both high and low temperature limits. In the high temperature limit, i.e. $T\gg m_i(\phi)$, we find~\cite{Arnold:1992rz} 
\begin{eqnarray} \label{hight T expansion for bosons}
V_T(\phi,T)&=&V_0( \phi) -\frac{\pi ^2}{90}T^4+\frac{1}{24}m_\phi^2(\phi )T^2-\frac{1}{12\pi}m_\phi^3(\phi )T+\mathcal{O} (m_\phi^4(\phi ))\nonumber\\
&&-\frac{7\pi ^2}{180}T^4+\frac{1}{12}m_f^2(\phi )T^2+\mathcal{O} (m_f^4(\phi ))\nonumber\\
&\approx&\frac{1}{6}g_{\phi f}T^2m_{f,0}\,\phi -\frac{1}{4}m_\phi^2\left( 1-\frac{T^2}{4\phi _{0}^{2}}-\frac{g_{\phi f}^2T^2}{3 m_{\phi}^2} \right) \phi ^2+...\,.
\end{eqnarray}
In the last line, we keep the field-dependent terms only at the leading order of the high temperature expansion. The corresponding $F$ and $G$ in Eq.~(\ref{eq:fullV2}) are then given by
\begin{eqnarray}\label{eq:FGT}
F_T=\frac{1}{6}g_{\phi f}T^2m_{f,0},\quad
G_T=\frac{1}{8}\frac{m_{\phi}^2}{\phi_0^2}T^2+\frac{1}{6} g_{\phi f}^2T^2\,.
\end{eqnarray}
Note that $F_T$ is linear in $g_{\phi f}$, whereas the contribution in $G_T$ is quadratic in the coupling. 
In the low temperature limit, i.e. $T\ll m_i(\phi)$, the finite-temperature corrections are exponentially suppressed and have negligible effects. 

Next, we consider the finite density effects by assuming zero temperature. If the cold Fermi gas of $\psi_f$ is part of the matter source for the stellar object, the Yukawa coupling term in Eq.~(\ref{eq:Lf}) gives rise to the finite-density correction,
\begin{eqnarray}
V_{\rho}(\phi)=g_{\phi f}\int \langle \bar{\psi}_f \psi_f \rangle \, d\phi\,,
\end{eqnarray}
where $\langle \bar{\psi}_f \psi_f \rangle$ denotes the field dependent number density of the Fermi gas, with
\begin{eqnarray} \label{Yukawa scalar}
\langle \bar{\psi}_f \psi_f \rangle &=&\frac{2}{(2\pi )^3}\int_0^{k_{F}^{}}{\mathrm{d}^3}p\frac{m_f(\phi)}{\sqrt{p^2+m_f^2(\phi)}}\nonumber\\
&=&\dfrac{1}{2\pi^2} m_f(\phi)\left( k_F\sqrt{m_{f}^{2}(\phi)+k_{F}^{2}}-m_{f}^{2}(\phi)\tanh ^{-1}\frac{k_F}{\sqrt{m_{f}^{2}(\phi)+k_{F}^{2}}} \right)\nonumber\\
 &\approx& \begin{cases}
	\dfrac{1}{2\pi ^2}k_{F}^{2}m_f\left( \phi \right) +\mathcal{O} (m_{f}^{3}),\text{\;}k_F\gtrsim m_f(\phi)\\[3mm]
	\dfrac{1}{3\pi ^2}k_{F}^{3}+\mathcal{O} (k_{F}^{4}),\text{\;}k_F\lesssim m_f(\phi)\\
\end{cases}.
\end{eqnarray}
where $k_F$ denotes the Fermi momentum at zero temperature. 
In the last line, we consider the high and low density limits to simplify the discussion. In the high density limit, i.e. $k_F\gg m_f(\phi)$, we obtain  $V_{\rho}(\phi)\approx \frac{1}{2\pi^2}k_F^2g_{\phi f}(m_{f,0}+\frac{1}{2}g_{\phi f}\phi)\phi$ at the leading order. In this case, the corresponding $F$ and $G$ are given by
\begin{eqnarray}\label{eq:FGrho1}
F_{\rho}\approx \frac{1}{2\pi^2}k_F^2g_{\phi f} m_{f,0}\,,\quad
G_{\rho}\approx\frac{1}{2\pi^2}k_F^2 g_{\phi f}^2\quad \textrm{(high $k_F$)}\,.
\end{eqnarray}
In the low density limit, i.e. $k_F\ll m_f(\phi)$, we obtain  $V_{\rho}(\phi)\approx\frac{1}{3\pi^2}g_{\phi f} k_F^3\phi$ at the leading order. Consequently, only the effective force term is present, resulting in
\begin{eqnarray}\label{eq:FGrho2}
F_{\rho}\approx \frac{1}{3\pi ^2} g_{\phi f} k_{F}^{3}=g_{\phi f} \, n_f,  \; (\text{low } k_F)
\end{eqnarray}
where $n_f$ denotes the number densities of fermions.

\section{Non-trivial scalar profiles for stellar objects}
\label{sec:stellar}

In this paper, we focus on the test field limit for the scalar and ignore its backreaction on the background spacetime (see the Appendix~\ref{app:backreaction} for a more detailed discussion of the backreaction of the scalar field).  
Additionally, we confine our discussion to a static, spherically symmetric and asymptotically flat spacetime, characterized by the following line element
\begin{equation} \label{metrics}
\mathrm{d}s^2=-B\left( r \right) \mathrm{d}t^2+A\left( r \right) \mathrm{d}r^2+r^2\mathrm{d}\theta ^2+r^2\sin ^2\theta \mathrm{d}\varphi ^2\,. 
\end{equation}
The equation of motion (EOM) for the test field, i.e. Klein-Gordon equation in the curved spacetime, is then given by
\begin{eqnarray}\label{eq:phiEOM0}
\frac{d^2\phi}{dr^2} +\left( \frac{2}{r}+\frac{\partial _rB}{2B}-\frac{\partial _rA}{2A} \right) \frac{d\phi}{dr}-A\frac{\partial V(\phi)}{\partial \phi}=0\,.
\end{eqnarray}
Here, $V(\phi)$ represents the full scalar potential in Eq.~(\ref{eq:fullV2}) with finite temperature or density corrections, and $T$ and $k_F$ in Eqs.~(\ref{eq:FGT}), (\ref{eq:FGrho1}), and (\ref{eq:FGrho2}) denote the proper temperature and Fermi momentum in the local inertia frame of the curved background. The scalar profile has to satisfy the appropriate boundary conditions at the origin and infinity, specifically  $\left.d\phi/dr\right|_{r=0}=0$ and $\phi(r\to\infty)=0$. 

In the following discussion, we compute the scalar profiles in ordinary stellar objects with hot and dense environments, as well as a candidate of not quite black holes with extremely high temperature and density in their high-curvature interior. The distinct behavior of metric functions and the matter sources in their interiors yield quite different predictions for the scalar charge. 
However, in the test field limit, the exteriors for all of these cases are well approximated by the vacuum solution. Therefore, let us first simplify the scalar field EOM at the exterior before discussing the interior for different cases.

It is useful to express the EOM in terms of dimensionless quantities. Specifically, we can rewrite the scalar field and metric as functions of the dimensionless radius $\bar{r}=r/R$, where $R$ is the radius of the object. The exterior spacetime can be then described by the rescaled metric below for any value of the total mass $M$: 
\begin{eqnarray}\label{eq:Schdmetric}
\bar{B}(\bar{r})&\equiv& B(r)\approx 1-\frac{C}{\bar{r}}\nonumber\\
\bar{A}(\bar{r})&\equiv& A(r)\approx
\left(1-\frac{C}{\bar{r}}\right)^{-1}\,,
\end{eqnarray}
where $C\equiv r_H/R\lesssim 1$ represents the dimensionless compactness, with $r_H=2M\lp^2$ being the horizon radius and $\lp$ being the Planck length. We can further normalize the scalar field by the VEV $\phi_0$ and define the rescaled scalar field $\varphi\equiv\phi/\phi_0$. Then, with $V(\phi)=V_0(\phi)$ in Eq.~(\ref{eq:phiEOM0}), the scalar field EOM at the exterior is simplified as
\begin{equation}\label{eq:EOMexterior}
   \frac{\mathrm{d}^2\varphi}{\mathrm{d}\bar{r}^2}+\left( \frac{2}{\bar{r}}+\frac{\partial _{\bar{r}}\bar{B}}{2\bar{B}}-\frac{\partial _{\bar{r}}\bar{A}}{2\bar{A}} \right) \frac{\mathrm{d}\varphi}{\mathrm{d}\bar{r}}-\bar{A}\left( \bar{r} \right) \left[ -\frac{1}{2}\eta^2 \varphi +\frac{1}{2}\eta^2\varphi ^3 \right] =0\,,
\end{equation} 
where $\eta\equiv m_\phi R$. At large distances, the metrics approach unity and the scalar field approaches the VEV. The equation in this limit then simplifies as $\partial^2_{\bar{r}}\delta\varphi+\frac{2}{\bar{r}}\partial_{\bar{r}}\delta\varphi-\frac{1}{2}\eta^2\delta\varphi=0$, where $\delta\varphi(\bar{r})\equiv \varphi(\bar{r})-1$ denotes the difference in the normalized field values with the VEV. The solution takes the Yukawa form with
\begin{eqnarray}\label{eq:deltaphiWeak}
\delta\varphi(\bar{r})
\approx \dfrac{1}{4\pi \bar{r}}\frac{Q}{\phi_0 R}e^{-\eta \bar{r}}\,,
\end{eqnarray}
where $\eta\lesssim 1$ ensures the mediation of a long-range force at the exterior. The scalar charge $Q$ can be obtained by matching Eq.~(\ref{eq:deltaphiWeak}) to the numerical solutions at a sufficiently large distance. 
To compare the strength of the scalar  force to that of gravity, it is useful to define a dimensionless scalar charge-to-mass ratio as follows:
\begin{equation} \label{gamma i}
\gamma =\frac{Q}{\sqrt{4\pi \lp^2} M}\,.
\end{equation}

\subsection{Ordinary stellar objects}
\label{sec:ordstellar}

For illustrative purposes, we consider the Sun (as a typical main sequence star), white dwarfs (WDs), and neutron stars (NSs) as examples of ordinary stellar objects that are either hot or dense. We approximate these objects as constant density stars as a good leading-order approximation to the more realistic solutions. The properties of these stellar objects are summarized in Table~\ref{Tab:stellar}.

\begin{table}[H]
\centering
\begin{tabular}{cccc}
\hline
                        & Sun                 & WDs      & NSs     \\ \hline
$M$ ($M_\odot$)        
& 1                   & 0.5 - 1.4         & 1.4-2.5             \\
$R$ (km)      
& $7\times10^{5}$ & $10^4$ &
10 - 15 \\
$T$ (keV)  
& $\sim 0.1$                & $\sim 10^{-2}$               & $\lesssim 10^5$            \\
$n_{N,e}$ (MeV$^3$) 
& $\sim 10^{-5}$    & $\sim 10^{-3}$        & $\sim 10^6$     \\ 

\hline
\end{tabular}
\caption{Physical properties of representative stellar objects~\cite{Lattimer:2015eaa, Siverd:2012fz, Steiner:2012xt}. $M$ is the total stellar mass in the unit of solar mass $M_\odot$. $R$ is the radius. $T$ denotes the average temperature. $n_{N}$ and $n_e$ denote the average number densities of nucleons and electrons, which are comparable in magnitude due to the charge neutrality condition. For NSs, $T$ denotes the highest temperature reached in a newly formed NS or a binary merger of NSs.
}
\label{Tab:stellar}
\end{table}

The interior metrics of a constant density star at $\bar{r}\lesssim 1$ can be solved exactly, with the rescaled counterparts as follows
\begin{eqnarray}\label{e2: reduced NS metric}
\bar{B}(\bar{r})=
	\dfrac{1}{4}\left( 3\sqrt{1-C}-\sqrt{1-C\,\bar{r}^2} \right) ^2,\quad
\bar{A}(\bar{r})=
	\left( 1-C\,\bar{r}^2 \right) ^{-1}\,.
\end{eqnarray}
Referring to Table~\ref{Tab:stellar}, we find that the compactness $C$ is approximately 0.3, $10^{-3}$ and $10^{-5}$ for NSs, WDs, and the Sun, respectively. 
By substituting the approximated form of $V(\phi)$ from Eq.~(\ref{eq:fullV2}) into Eq.~(\ref{eq:phiEOM0}), the EOM for the rescaled scalar field in the interior, i.e. $\bar{r}\leq1$, is simplified as 
\begin{equation}
   \frac{\mathrm{d}^2\varphi}{\mathrm{d}\bar{r}^2}+\left( \frac{2}{\bar{r}}+\frac{\partial _{\bar{r}}\bar{B}}{2\bar{B}}-\frac{\partial _{\bar{r}}\bar{A}}{2\bar{A}} \right) \frac{\mathrm{d}\varphi}{\mathrm{d}\bar{r}}-\bar{A}\left( \bar{r} \right) \left[ \bar{F}+\left(\bar{G} -\frac{1}{2}\eta^2 \right) \varphi +\frac{1}{2}\eta^2\varphi ^3 \right] =0\,,
\end{equation} 
where $\bar{F}=F R^2/\phi_0$ and $\bar{G}=G R^2$ denote the corresponding dimensionless quantities. This equation can be directly matched to the exterior EOM given in Eq.~(\ref{eq:EOMexterior}) at $\bar{r}=1$. 

For a stellar object with a specific value of $\bar{F}$ or $\bar{G}$, the non-trivial scalar profile can be numerically solved using the shooting method. By imposing the boundary condition $\left.d\varphi/d \bar{r}\right|_{\bar{r}=0}=0$ 
at the origin, the scalar field value at the origin, i.e. $\delta \varphi_0\equiv \varphi(\bar{r}=0)-1$, can be determined by requiring the solution to decay asymptotically at infinity, i.e., $\varphi(\bar{r}\to \infty)=0$. This also determines the field value at the boundary, i.e. $\delta \varphi_1\equiv \varphi(\bar{r}=1)-1$, as a function of $\bar{G}$ or $\bar{F}$. 
For later discussion, we define the deviation of the rescaled scalar field from the VEV as follows:
\begin{eqnarray}
  \delta\varphi(\bar{r})\equiv \varphi(\bar{r})-1\,.
\end{eqnarray}
Since the radius $R$ of ordinary stellar objects is much larger than $r_H$, the scalar charge can be obtained by matching the weak gravity expansion in Eq.~(\ref{eq:deltaphiWeak}) with the numerical solution at $\bar{r}=1$. This yields $Q\approx 4\pi \delta\varphi_1\phi_0 R$, and the value of the scalar charge-to-mass ratio $\gamma$ is determined from Eq.~(\ref{gamma i}), i.e.
\begin{equation} \label{gamma i1}
\gamma \approx 4\sqrt{\pi}\,C^{-1}\delta\varphi_1\, \frac{\phi _0}{m_{\rm pl}}\,,
\end{equation}
which relies on the properties of the scalar field as well as the stellar objects.

\begin{figure}[!h]
	\centering
	\includegraphics[height=5.1cm]{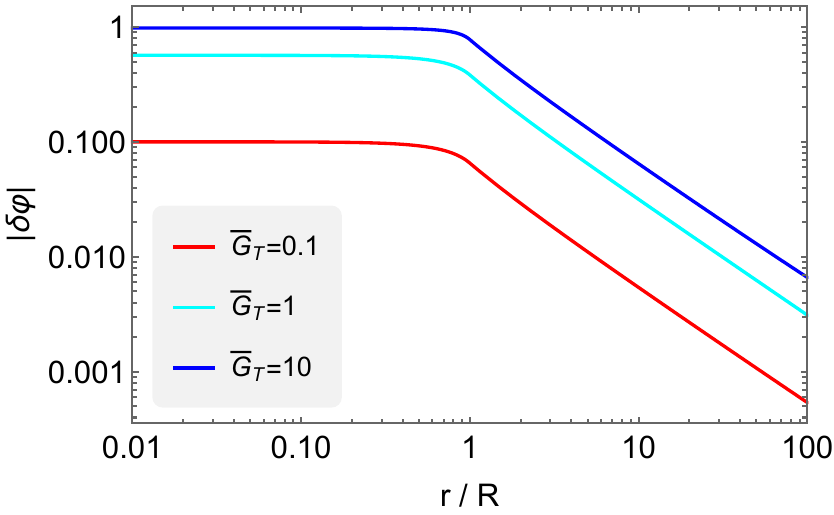}\;\;
	\includegraphics[height=5.1cm]{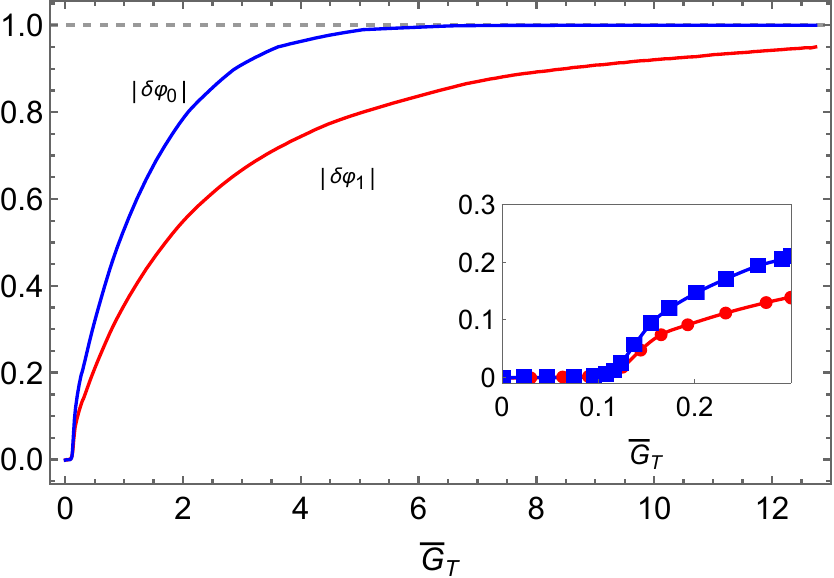}	
    \caption{Finite temperature effects induced by the scalar self-interaction in ordinary stellar objects. Left: the rescaled scalar profile $\delta\varphi(\bar{r})$ as a function of the rescaled radius $\bar{r}=r/R$ for several benchmark values of the dimensionless coefficient $\bar{G}_T$. Right:  the rescaled scalar field values $|\delta\varphi_0|$ and $|\delta\varphi_1|$ as functions of $\bar{G}_{T}$, when the condition $\bar{G}_T\gtrsim \eta^2/2$ is satisfied.}
    \label{fig:OSG}
  \end{figure}

Now, we will discuss the environment effects inside ordinary stellar objects for various cases. 
Let us first consider the finite temperature effects induced by the self-interaction of the scalar field. Referring to Table~\ref{Tab:stellar}, we observe that the scalar mass allowing for a long-range force ($\eta\lesssim 1$) for the three stellar objects are significantly smaller than the average temperature $T$. Therefore, we can consider the high temperature limit, where the corrections manifest as the 
$G$-term in Eq.~(\ref{eq:FGT}). Its dimensionless counterpart is given by
\begin{eqnarray}
\bar{G}_T=\frac{\eta^2}{8}\frac{T^2}{\phi_0^2}\,.
\end{eqnarray}
As demonstrated in Fig.~\ref{fig:VphiFG} and explained below Eq.~(\ref{eq:fullV2}), a change in sign for the quadratic term, i.e., $\bar{G}_T\gtrsim\eta^2/2$, is required to ensure a non-trivial scalar profile for this case. This sets an upper bound on $\phi_0$, given by $\phi_0\lesssim 2T$. Once this condition is met, we can solve for the scalar profile for a given value of $\bar{G}_T$, where there is a one-to-one mapping between $\bar{G}_T$ and the negative field values of $\delta\varphi_0$ and $\delta\varphi_1$.

The left panel of Fig.~\ref{fig:OSG} shows the rescaled scalar profiles for several benchmark values of the dimensionless coefficient $\bar{G}_T$, as obtained from numerical solutions. The field value exhibits mild variation within the stellar objects and follows a $1/r$ decay outside, before the onset of mass suppression. The right panel shows $|\delta\varphi_0|$ and $|\delta\varphi_1|$ as functions of $\bar{G}_T$.
As illustrated in the inset, a non-trivial scalar profile exists only when the corrections surpass a certain threshold, namely $\bar{G}_T\gtrsim 0.1$ for this case. This is similar to that observed for the QCD axion within neutron stars~\cite{Hook:2017psm}, where the exact threshold value varies depending on the shape of scalar potential.
Above the threshold, the magnitude of $|\delta\varphi_0|$ increases linearly with $\bar{G}_T$ for small $\bar{G}_T$ and approaches unity in the limit of large $\bar{G}_T$, corresponding to the boundary value moving asymptotically to the new maximum of $-V(\phi)$ at $\phi=0$. The magnitude of $|\delta\varphi_1|$ remains slightly smaller than that of $|\delta\varphi_0|$, with their difference diminishing as $\bar{G}_T$ increases.

\begin{figure}[!h]
    \centering
    \includegraphics[height=6cm]{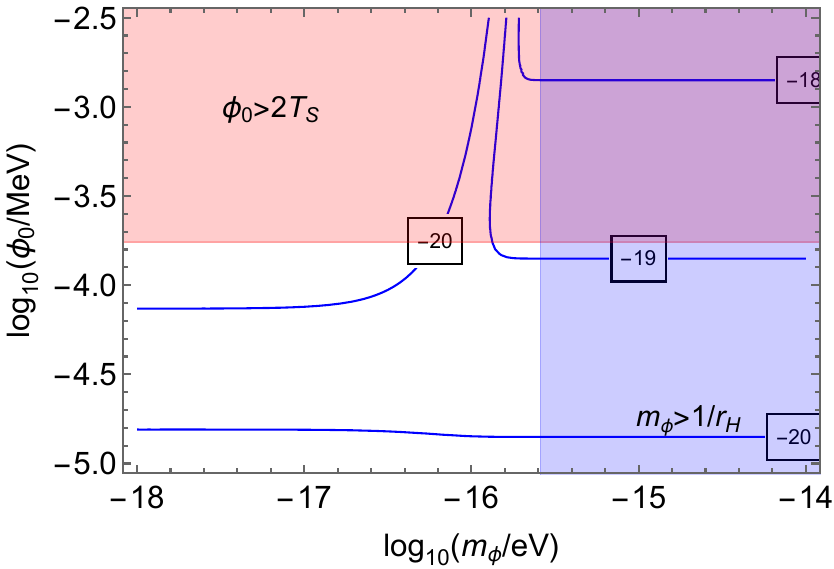}
    \caption{Contours of the scalar charge-to-mass ratio $|\gamma|$ for the Sun on the scalar mass $m_{\phi}$ and VEV $\phi_0$ plane. The horizontal line denotes the upper bound $\phi_0\lesssim 2T$. The vertical line denotes the condition of a long-range force, i.e. $\eta\lesssim 1$.}
    \label{fig:contourSG}
\end{figure}

Figure~\ref{fig:contourSG} presents the contours of the scalar charge-to-mass ratio $|\gamma|$ given in Eq.~(\ref{gamma i1}) for the Sun in the plane of scalar mass and the VEV. The absolute value of $\gamma$ is strongly suppressed, with $|\gamma|\lesssim 10^{-19}$, due to the upper bound on $\phi_0$ set by $T$. Similar results are obtained for WDs and NSs, indicating that the finite temperature effects with only the scalar self-interaction are completely negligible.

Next, we consider the Yukawa couplings of the scalar field to either electrons or nucleons. Because of the strong constraints on these couplings, the scalar field couples weakly to the SM fermions and makes negligible contribution to their mass. Referring to Table~\ref{Tab:stellar}, we find that all three examples fall within the low temperature or density limits, i.e. $m_e, \, m_N\gg T$ or $m_N\gg n_N^{1/3}$. Therefore, the dominant corrections come from the finite density effects, characterized by the $F$-term in Eq.~(\ref{eq:FGrho2}). Its  dimensionless counterpart is given by
\begin{eqnarray}
\bar{F}_\rho=g_{\phi f} \frac{n_f R^2}{\phi_0}\,.
\end{eqnarray}
As shown in Fig.~\ref{fig:VphiFG} and explained below Eq.~(\ref{eq:fullV2}), a non-trivial scalar profile can be found when $\bar{F}_{\rho}$ is negative, i.e. $g_{\phi f}<0$, with the corresponding  $\delta\varphi_0$ and $\delta\varphi_1$ being positive. 

\begin{figure}[H]
    \begin{minipage}{0.5\textwidth}
      \centering
       \includegraphics[height=5.1cm]{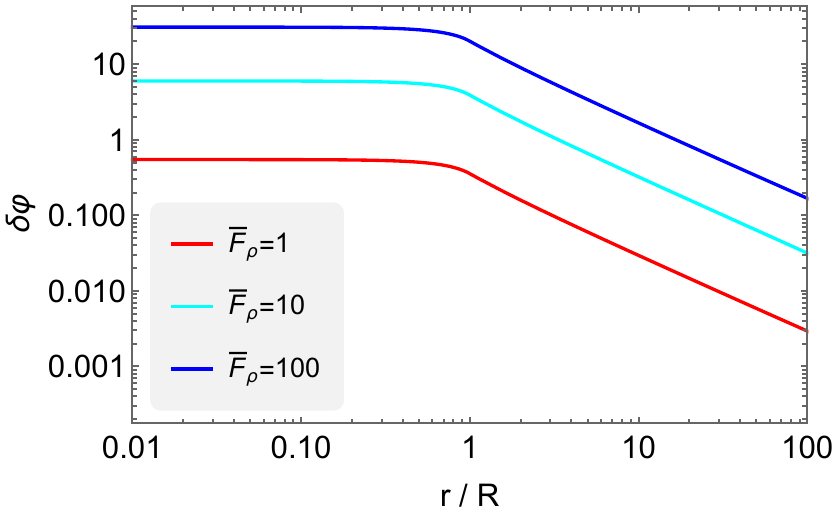}
      \end{minipage}
 \begin{minipage}{0.5\textwidth}
      \centering
       \includegraphics[height=5.1cm]{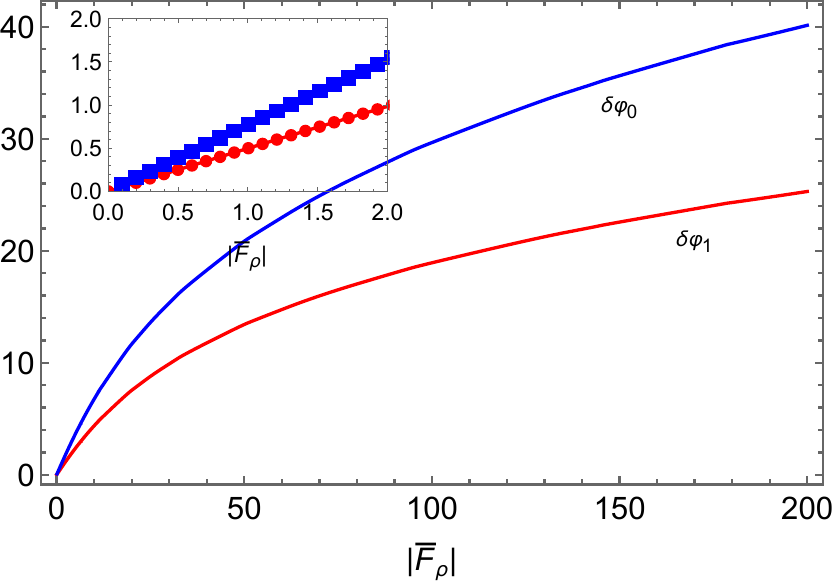}
      \end{minipage}
        \caption{Finite density effects induced by the Yukawa couplings to the SM fermions in ordinary stellar objects. Left: the rescaled scalar profile $\delta\varphi(\bar{r})$ as a function of the rescaled radius $\bar{r}$ for several benchmark values of the dimensionless coefficient $\bar{F}_{\rho}$. Right:  the rescaled scalar field values $\delta\varphi_0$ and $\delta\varphi_1$ as functions of $|\bar{F}_\rho|$. }
        \label{fig:OSF}
  \end{figure}

The left panel of Fig.~\ref{fig:OSF} shows the rescaled scalar profiles for several benchmark values of the dimensionless $\bar{F}_{\rho}$. The right panel presents the numerical solutions for $\delta\varphi_0$ and $\delta\varphi_1$ as functions of $|\bar{F}_\rho|$. Unlike the $G$-term case, we note the absence of a threshold required for the existence of non-trivial scalar profiles, as demonstrated in the inset. Moreover, both $\delta\varphi_0$ and $\delta\varphi_1$ continue to increase as $|\bar{F}_\rho|$ becomes large. This is due to the new maximum of $-V(\phi)$, as shown in Fig.~\ref{fig:VphiFG}, which consistently shifts to larger values with increasing $|\bar{F}_\rho|$, theoretically allowing for a larger value of the scalar charge. The ratio $\delta\varphi_1/\delta\varphi_1$ has a smaller value compared to the $G$-term case, and it decreases slowly as $|\bar{F}_\rho|$ increases.

\begin{figure}[!h]
    \centering
    \includegraphics[height=6cm]{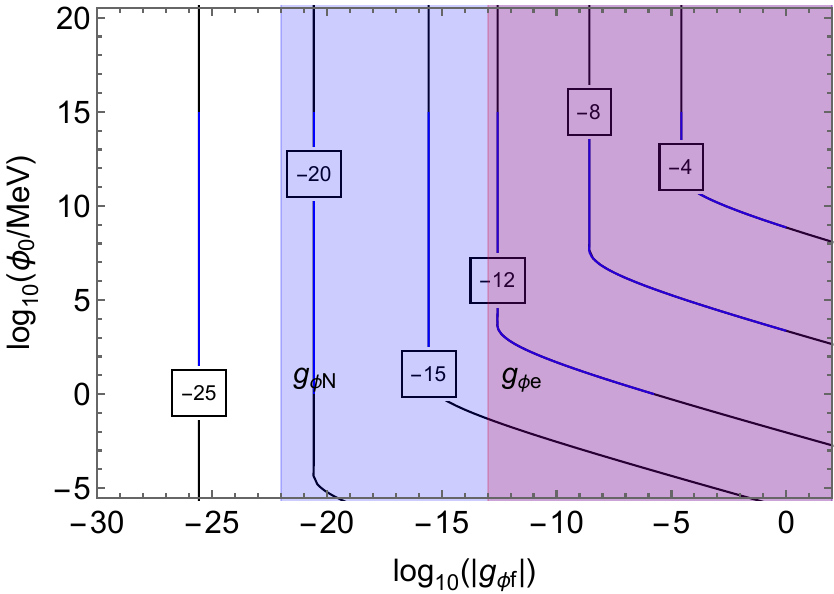}
    \caption{Contours of the scalar charge-to-mass ratio $\gamma$ for the WDs on the scalar field VEV $\phi_0$ and the Yukawa coupling $g_{\phi f}$ ($f=e,\,N$) plane. The blue and red vertical lines denote upper limits on $g_{\phi N}$ \cite{KumarPoddar:2020kdz} and $g_{\phi e}$ \cite{Viaux:2013lha}, respectively.}
    \label{fig:contourSF}
\end{figure}

Figure~\ref{fig:contourSF} displays the contours of $\gamma$ for WDs in the plane of scalar VEV and the Yukawa couplings to either electrons or nucleons. Because $n_e$ and $n_N$ are of the same magnitude due to charge neutrality, we assume $n_e\approx n_N$ here, and the results for $g_{\phi e}$ and $g_{\phi N}$ are the same.
When $\phi_0$ is sufficiently large, $|\bar{F}_\rho|$ is small, and $\delta\varphi_0$ increases linearly with its magnitude. Thus, the contour of $\gamma$ becomes independent of $\phi_0$, with $\gamma\sim 2\sqrt{\pi}C^{-1}g_{\phi f}n_f R^{2}/m_{\rm pl}$. However, for small $\phi_0$ and large $|\bar{F}_\rho|$, the increase in $\delta\varphi_0$ with $|\bar{F}_\rho|$ becomes slower, leading to a growth in $\gamma$ as $\phi_0$ increases.
Among the two cases considered, a larger $\gamma$ can be obtained if the scalar field couples to electrons, due to the weaker limits on $g_{\phi e}$.  Specifically, we obtain $\gamma\lesssim 10^{-12}$ for this case, which is considerably larger than the value obtained from the finite temperature effects. The value of $\gamma$ is smaller for either the Sun or NSs due to their smaller density or smaller size of the objects.

Therefore, in the case of the minimal model of the scalar field considered in Sec.~\ref{sec:scalar}, the environmental effects found in typical ordinary stellar objects are unable to generate a significant scalar charge. Specifically, considering the various constraints, the maximum achievable absolute value  of the scalar charge-to-mass ratio $\gamma$ cannot exceed $10^{-12}$, which naturally evades the stringent constraints from the fifth-force searches in laboratories or in astronomical observations. To probe the light scalar field for the minimal model, it is therefore necessary to explore more extreme environments, as we will discuss in the following subsection.

\subsection{Not quite black holes: 2-2-holes}  \label{sec: 2-2-holes}

An intriguing candidate for UCOs is 2-2-holes~\cite{Holdom:2002xy, Holdom:2016nek}, which represent a novel class of solutions of the classical action described below:
\begin{eqnarray}\label{eq:CQG}
S_{\mathrm{CQG}}=\frac{1}{16\pi}\int d^4x\,\sqrt{-g}\left(\Mp^2R-\alpha C_{\mu\nu\alpha\beta}C^{\mu\nu\alpha\beta}+\beta R^2\right),
\end{eqnarray}
where $R$ is the Ricci scalar and $C_{\mu\nu\rho\sigma}$ is the Weyl tensor. Here, $\Mp$ denotes the Planck mass and $\alpha, \beta\gtrsim 1$ are dimensionless couplings associated with the quadratic curvature terms. 
Instead of being viewed as a truncation of the effective field theory for gravity at low energy, Eq.~(\ref{eq:CQG}) is considered a classical approximation of the renormalizable and asymptotically free quantum quadratic gravity~\cite{Stelle:1976gc}. Specifically, it is dominated by the quadratic curvature terms at high energy, while reducing to GR at low energy. This framework allows for the description of solutions encompassing both low and high curvature regimes~\cite{Holdom:2016nek}.

The existence of 2-2-holes critically depends on the Weyl term $C^{\mu\nu\rho\sigma}C_{\mu\nu\rho\sigma}$. In the presence of a compact matter source, such as a photon gas or cold Fermi gas, it has been found that horizonless 2-2-hole solutions exist with an arbitrary mass $M$ above the minimal value $\Mmin\sim \Mp^2 \lambda_2$~\cite{Ren:2019afg,Aydemir:2021dan}. Here, the parameter $\lambda_2\sim \sqrt{\alpha}\,\lp$ represents the Compton wavelength of the new spin-2 mode associated with the Weyl term.\footnote{The Weyl term brings in the problematic spin-2 ghost with mass $m_2=1/\lambda_2$ in the classical theory. However, the fate of the ghost remains under debate at the quantum level. See Ref.~\cite{Salvio:2018crh} for a review on quadratic gravity and references therein.} In quadratic gravity, 2-2-holes appear to be a more general class of solutions than black hole solutions. Therefore, it is highly likely that 2-2-holes serve as the end points of gravitational collapse in this theory~\cite{Holdom:2016nek}.

\begin{figure}[!h]
  \centering%
{ \includegraphics[width=8cm]{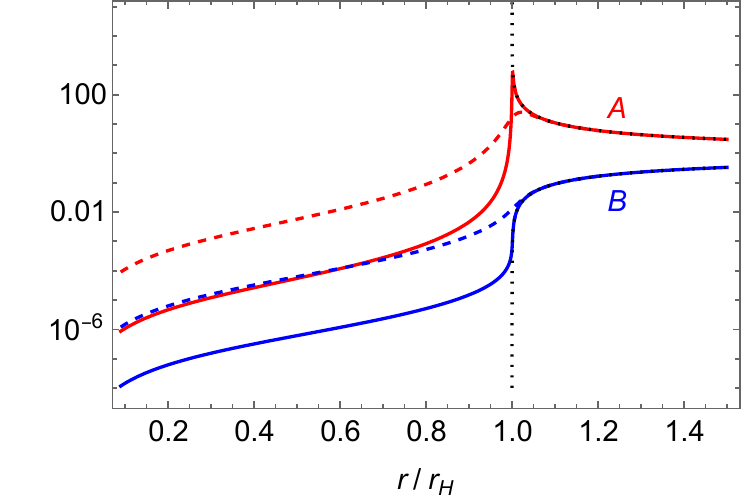}}\;
{ \includegraphics[width=8cm]{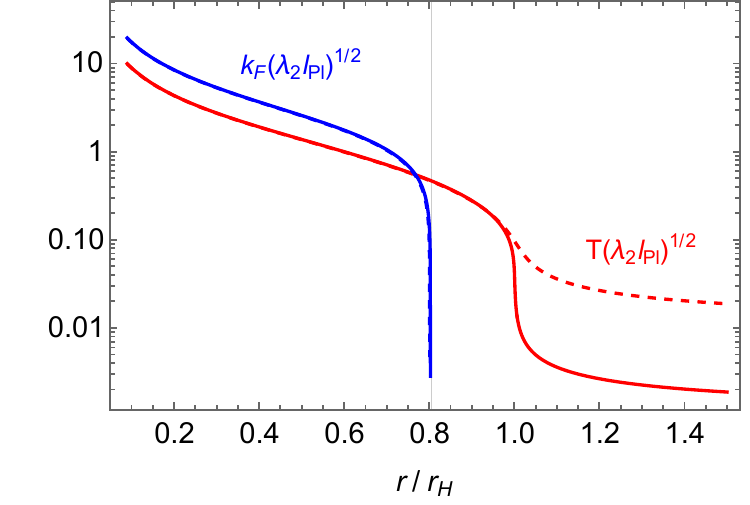}}
\caption{\label{fig:22hole} 
Properties of 2-2-holes sourced by a photon gas or cold Fermi gas. Left: the metric $A$ (red) and $B$ (blue) as functions of the rescaled radius $\bar{r}=r/r_H$. The black dotted lines denote the Schwarzchild solution. Right: the proper temperature $T$ for the photon gas (red) and the proper Fermi momentum $k_F$ for the cold Fermi gas with mass $m_f(\lambda_2\lp)^{1/2}=1$ (blue) as functions of $\bar{r}$. The vertical gray line denotes the radius $r_F$ where $k_F$ drops quickly to zero. In both panels, the dashed and solid lines denote solutions with $r_H/\lambda_2\approx 10$ and $100$, respectively. }
\end{figure}

A typical 2-2-hole with a mass $M\gg M_{\rm min}$ displays distinctive behaviors~\cite{Ren:2019afg,Aydemir:2021dan}, as illustrated in Fig.~\ref{fig:22hole}. Beyond the would-be horizon at $r_H$, the matter density becomes negligible, and the 2-2-hole metrics closely resemble those of a black hole with the same mass, i.e.
\begin{eqnarray}
 A(r)\approx\frac{1}{B(r)}\approx\left(1-\frac{r_H}{r}\right)^{-1}\,,
\end{eqnarray}
due to the dominance of the Einstein term.
There exists a narrow transition region at $r\sim r_H$, where the metric functions begin to significantly deviate from the black hole solutions at a small distance just outside $r_H$. In this region, the quadratic curvature terms in Eq.~(\ref{eq:CQG}) start to compete with the Einstein term, making the derivation of an analytical solution challenging.
At $r\lesssim r_H$, the quadratic curvature terms become dominant, leading to an extremely high curvature region in the interior. As the distance $r$ decreases, the metrics $A(r)$ and $B(r)$ approach zero following a $r^2$ dependence, indicating the presence of a timelike singularity at the origin. 
In the small $r$ region, the metric functions can be closely approximated by series expansion, with~\cite{Ren:2019afg}
\begin{eqnarray}
  \begin{aligned}
    A\left( r \right) \approx a_2r^2\left[ 1+4\sqrt{a_2\Mp^2}r^2+\frac{27}{2}a_2\Mp^2r^4+\mathcal{O} \left( r^6 \right) \right], 
\\
B\left( r \right) \approx b_2r^2\left[ 1+3\sqrt{a_2\Mp^2}r^2+\frac{15}{2}a_2\Mp^2r^4+\mathcal{O} \left( r^6 \right) \right],
  \end{aligned}
\end{eqnarray}
where $a_2$ has a one-to-one mapping to the mass $M$, and $b_2$ is determined by the normalization of $B(r)$ at spatial infinity. 

A small value of $B(r)$ in the interior also indicates a deep gravitational potential. By applying the momentum conservation law to the stress tensor, the gases satisfy the generalized versions of Tolman’s law, namely $T^2(r)B(r)$ and $(k_F^2(r)+m_f^2)B(r)$ remain constants for the photon gas and cold Fermi gas, respectively~\cite{Ren:2019afg,Aydemir:2021dan}. Consequently, as one approaches the origin of the 2-2-hole, the gas would exhibit either high temperature or density, attributed to the decreasing $B(r)$.
This creates an extreme environment capable of sourcing weakly coupled scalar fields and may provide access to new physics that would otherwise remain inaccessible. Specifically, the photon gas and cold Fermi gas exhibit similar behavior in the deep interior, characterized by exceptionally high temperature or Fermi momentum, with $T(r), k_F(r)\propto 1/r$. The difference becomes evident at larger radii. The Fermi momentum $k_F$ rapidly decreases to zero within the interior at $r_F$, when it becomes comparable to the mass $m_f$. In contrast, the temperature $T$ for the photon gas extends to the would-be horizon and decreases significantly only in the exterior.

In investigating the non-trivial scalar profile created by astrophysical 2-2-holes, where $r_H$ is significantly larger than $\lambda_2$,  it is advantageous to make certain approximations to the numerical 2-2-hole solutions. As the mass $M$ increases, a 2-2-hole progressively resembles a black hole from the exterior. Consequently, it is a reasonable approximation to utilize the EOM in  Eq.~(\ref{eq:EOMexterior}) for the exterior, along with Eq.~(\ref{eq:Schdmetric}) and $R\approx r_H$. On the other hand, the metrics and matter properties in the interior exhibit a novel scaling behavior with $M$ (or $r_H$) due to the dominance of quadratic curvature terms, particularly in the limit of $M\gg M_{\rm min}$. Specifically, at the leading order of high curvature expansion, the solutions can be fully characterized by the following dimensionless quantities~\cite{Ren:2019afg,Aydemir:2021dan}, which are functions of the rescaled radius $\bar{r}=r/r_H$,
\begin{eqnarray}\label{eq:22holescaling}
\bar{A}(\bar{r})=A(r)\frac{r^2_{H}}{\lambda^2_{2}},\quad
\bar{B}(\bar{r})=B(r)\frac{r^2_{H}}{\lambda^2_{2}},\quad
\bar{T}(\bar{r})=T(r)\sqrt{\lambda_2\lp},\quad
\bar{k}_F(\bar{r})=k_F(r)\sqrt{\lambda_2\lp}\,.
\end{eqnarray}
A simplified scalar EOM for the normalized field $\varphi(r)$ in the 2-2-hole interior is then obtained by  substituting the full potential $V(\phi)$ from Eq.~(\ref{eq:fullV2}) and the scaling behavior from Eq.~(\ref{eq:22holescaling}) into Eq.~(\ref{eq:phiEOM0}), 
\begin{equation}\label{eq:rescaleEOM22hole}
   \frac{\mathrm{d}^2\varphi}{\mathrm{d}\bar{r}^2}+\left( \frac{2}{\bar{r}}+\frac{\partial _{\bar{r}}\bar{B}}{2\bar{B}}-\frac{\partial _{\bar{r}}\bar{A}}{2\bar{A}} \right) \frac{\mathrm{d}\varphi}{\mathrm{d}\bar{r}}-\bar{A}\left( \bar{r} \right) \left[ \bar{F}+\left(\bar{G} -\frac{1}{2}\zeta^2 \right) \varphi +\frac{1}{2}\zeta^2\varphi ^3 \right] =0\,,
\end{equation} 
where $\bar{F}=F \lambda_2^2/\phi_0$, $\bar{G}=G \lambda_2^2$ and $\zeta\equiv m_\phi \lambda_2\lesssim \eta$. 
In contrast to ordinary stellar objects, the dimensionless quantities in the 2-2-hole interior scale with the intrinsic length scale $\lambda_2$ of quadratic gravity, rather than the physical size of the object. This leads to distinct predictions regarding the scalar charge, as we will explore below.

Another point we would like to highlight for solving scalar profiles for 2-2-holes is the appropriate choice of boundary condition for the scalar field at the origin. This point is characterized by a timelike curvature singularity, in contrast to the regular spacetime associated with ordinary stellar objects. Although the singularity at the origin might lead to geodesic incompleteness, it does not necessarily indicate a genuine physical ambiguity. More explicitly, from the perspective of relativistic classical field theories,  the field dynamics on a singular and so-called inextendible spacetime can be defined by adopting the mathematical framework proposed by Wald~\cite{Wald:1980jn}. Intuitively, this approach involves disregarding solutions with diverging energy as $r$ approaches the origin. If among the two linearly independent solutions, $\varphi_1(\bar{r})$ and $\varphi_2(\bar{r})$, a unique solution remains, the classical wave equation is considered well defined, and the singularity introduces no ambiguity. For more comprehensive details, please refer to Sec.~IIIC in Ref.~\cite{Holdom:2016nek}. This scenario is indeed applicable to 2-2-holes. With $A(r), B(r)\propto r^2$ in the small $r$ limit, we find $\varphi_1(\bar{r})\propto \bar{r}^0$ and $\varphi_2(\bar{r})\propto \bar{r}^{-1}$. Only $\varphi_{1}(\bar{r})$ has finite energy, thus necessitating a Neumann boundary condition for the scalar field at the origin. That is, $\left.d\varphi_{1}(\bar{r})/d \bar{r}\right|_{\bar{r}=0}=0$.

As $M$ increases, the boundaries of both the interior and exterior scaling regions shift toward $r_H$. This results in a transition region with a decreasing radial size but more significant spatial variations. Obtaining the exact solution of the transition region is challenging due to the limitations of numerical accuracy. Therefore, for our numerical study of the scalar profiles in this work, we choose to disregard the contribution from the narrow transition region around $r_H$. Instead, we directly match the rescaled EOM in Eq.~(\ref{eq:rescaleEOM22hole}) for the interior to that in Eq.~(\ref{eq:EOMexterior}) for the exterior at $\bar{r}\sim 1$. From our numerical solutions with $r_H/\lambda_2\lesssim \mathcal{O}(100)$, we have verified that the contribution from the transition region is indeed negligible for our order of magnitude estimation of the scalar charges of 2-2-holes.
Under this approximation, the rescaled scalar profile $\varphi(\bar{r})$ becomes independent of the 2-2-hole size. 
The scalar charge is again obtained by matching the numerical solution to the weak gravity expansion in Eq.~(\ref{eq:deltaphiWeak}) at some $\bar{r}_0\gg1$. This yields the scalar charge $Q\approx \pi \delta\varphi_1\phi_0 r_H$, and the scalar charge-to-mass ratio
\begin{equation} \label{gamma i2}
\gamma \approx \sqrt{\pi}\,C^{-1}\delta\varphi_1\, \frac{\phi _0}{\Mp}\,.
\end{equation}
In comparison to Eq.~(\ref{gamma i1}), where the weak gravity expansion is applicable at the surface of ordinary stellar objects, the magnitude of $\gamma$ for 2-2-holes is suppressed by approximately a factor of 4 due to the significant redshift around $r_H$.

Note that the approximated value of $\gamma$ in Eq.~(\ref{gamma i2}) is independent of $M$, making it universal for 2-2-holes of all masses $M\gg M_{\rm min}$. This is in contrast to the behavior observed in ordinary stellar objects discussed in Sec.~\ref{sec:ordstellar}, as well as in cases of spontaneous scalarization in scalar-tensor theories. In the latter, it has been argued that smaller black holes will be more strongly charged due to their larger curvature near the horizon~\cite{Maselli:2020zgv}.
If we take into account the contribution from the transition region, we may expect a $\gamma$ with a mild dependence on $M$.
Moreover, while the gas profiles are influenced by the particle mass, the scalar charges of 2-2-holes exhibit very little dependence on the mass, indicating that the scalar properties are primarily determined by the high curvature region in deep interior.

Now, let us consider several examples to illustrate the environmental effects within the framework of 2-2-holes.
We first examine the finite temperature effects arising from scalar self-interaction. For a 2-2-hole sourced by a photon gas, the temperature grows large in the interior for decreasing $r$. Consequently, we can take the high temperature limit and the corrections take the form of a $G_T$-term. The corresponding dimensionless counterpart can be expressed as:
\begin{eqnarray}
\bar{G}_T(\bar{r})=\frac{1}{8}\frac{m_{\phi}^2}{\phi_0^2}T^2(r)\lambda_2^2
=\frac{\bar{\lambda}_2}{8}\frac{m_{\phi}^2}{\phi_0^2}\bar{T}^2(\bar{r})
\equiv \mathcal{G}_T\bar{T}^2(r) \,,
\end{eqnarray}
where $\bar{T}(\bar{r})$ is the rescaled proper temperature as shown in Fig.~\ref{fig:22hole}, and $\mathcal{G}_T\equiv \bar{\lambda}_2m_{\phi}^2/(8\phi_0^2)$ represents the $r$-independent dimensionless coefficient. The condition for the sign flip of the quadratic term is given by $\phi_0\lesssim \bar{T}(\bar{r})\Mp/\sqrt{\bar{\lambda}_2}$. As $\bar{T}(\bar{r})$ continues to increase for smaller values of $\bar{r}$, for any given value of $\phi_0$, there will always be a radius below which this condition is satisfied. This implies that, unlike ordinary stellar objects where the magnitude of $T$ is limited within the stars, this condition can be easily fulfilled within the interior of 2-2-holes.

\begin{figure}[!h]
	\centering
	\includegraphics[height=5.1cm]{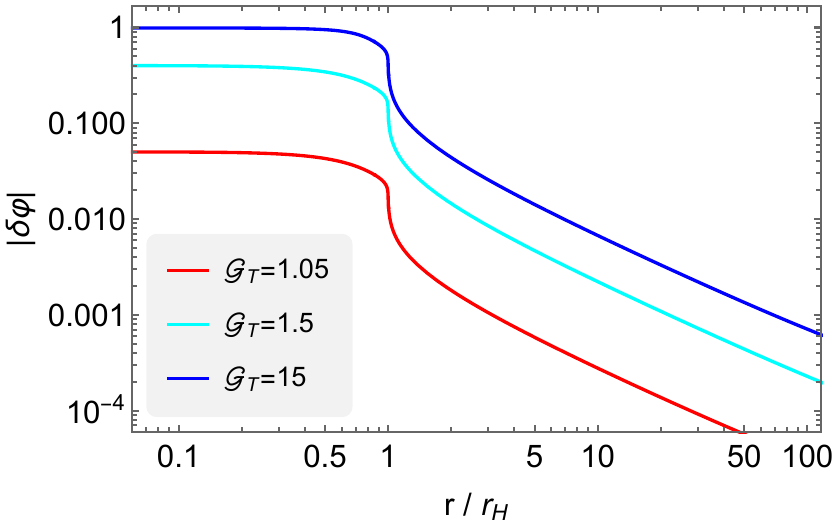}\;\;
	\includegraphics[height=5.1cm]{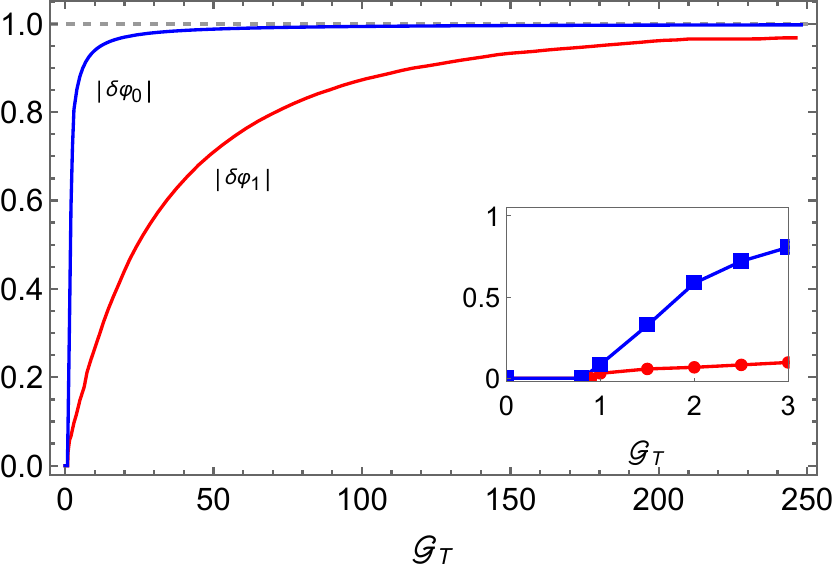}
\caption{\label{fig:22holeG} {Finite temperature effects induced by the scalar self-interaction for 2-2-holes. Left: the rescaled scalar profile $\left|\delta\varphi\right|$ as a function of the rescaled radial coordinate $\bar{r}=r/r_H$ for several benchmark values of the dimensionless coefficient $\mathcal{G}_T$. 
Right: rescaled scalar field values $|\delta\varphi_0|$ and $|\delta\varphi_1|$ as functions of $\mathcal{G}_{T}$.}
}
\end{figure}

\begin{figure}
    \centering
    \includegraphics[height=6cm]{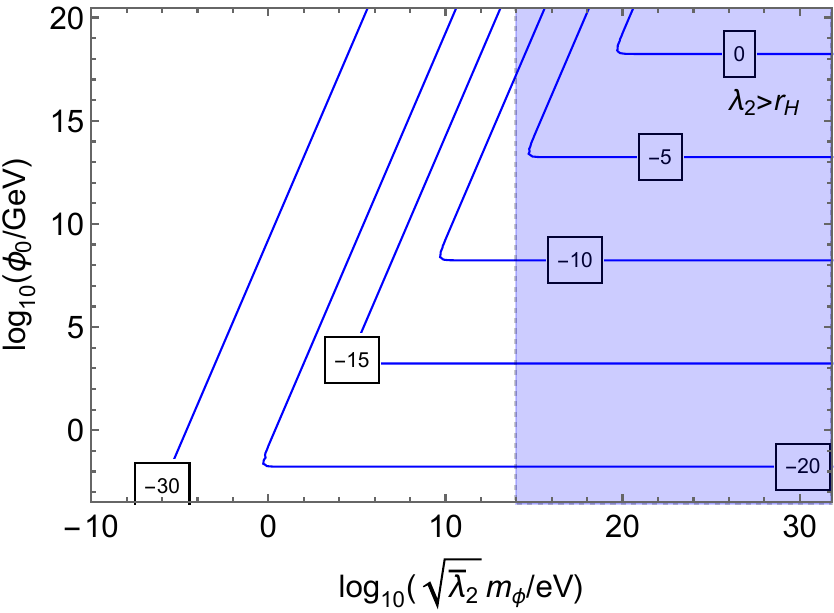}
    \caption{Contours of the scalar charge-to-mass ratio $|\gamma|$ on the plane of the scalar VEV $\phi_0$ and the combination $\bar{\lambda}_2 m_{\phi}$. The vertical line denotes the upper bound due to the requirement $\lambda_2\lesssim r_H$ and $\eta\lesssim 1$ for primordial 2-2-holes with $M\approx 10^{-10}\,M_\odot$. }
    \label{fig:contour2G}
\end{figure}

In the regime where $\bar{G}_T(\bar{r})\gg \zeta^2/2$, there is a one-to-one mapping between $\mathcal{G}_T$ and $\delta\varphi_0$ ($\delta\varphi_1$) for the numerical solutions. 
The left panel of Fig.~\ref{fig:22holeG} shows the rescaled scalar profiles for several benchmark values of the dimensionless $\mathcal{G}_T$. The right panel illustrates the behavior of $|\delta\varphi_0|$ and $|\delta\varphi_1|$ as functions of $\mathcal{G}_T$. Despite the differences in the $\bar{r}$-dependent terms, such as $\bar{A}(\bar{r})$ and $\bar{T}(\bar{r})$, the general behavior of the scalar profiles for 2-2-holes in Fig.~\ref{fig:22holeG} is similar to that for ordinary stellar objects shown in Fig.~\ref{fig:OSG}. However, due to the scaling behavior transition in 2-2-holes, $|\delta\varphi(\bar{r})|$ experiences a more pronounced drop around $\bar{r}\sim 1$, resulting in a greater suppression of $|\delta\varphi_1|$ compared to $|\delta\varphi_0|$ for 2-2-holes. Also, the threshold value of $\mathcal{G}_T$ required for the existence of a non-trivial scalar profile is different, namely $\mathcal{G}_T\gtrsim 1$ for the same model.

The contours of $\gamma$ in Fig.~\ref{fig:contour2G} exhibit a similar shape as that in Fig.~\ref{fig:contourSG}. The difference in magnitude between the two cases is determined by the ratio $\mathcal{G}_T/\bar{G}_T\approx \sqrt{\bar\lambda_2}/(TR)^2$, which reflects their different scaling behaviors. 
For 2-2-holes, the radial size of the interior shrinks to the order of $\lambda_2$ due to high curvature effects, limiting the temperature and radius product to $\sqrt{\bar{\lambda}_2}$, which is significantly smaller than the product $T R \sim 10^{17}$ for the Sun.
Furthermore, with $\lambda_2\lesssim r_H$ and $\eta\lesssim 1$, there is an upper bound on the horizontal axis, specifically $\sqrt{\bar{\lambda}_2}m_\phi\lesssim \sqrt{\bar{\lambda}_2}/r_H\lesssim 10^{9}\,\textrm{eV}\sqrt{M_{\odot}/M}$. As a result, the magnitude of $\gamma$ for astrophysical 2-2-holes with $M\gtrsim M_{\odot}$ is even more strongly suppressed compared to the Sun.
However, for primordial 2-2-holes with $M\ll M_{\odot}$, a larger $|\gamma|$ can be achieved due to the increasing allowed range of $m_\phi$, corresponding to a larger self-interaction coupling.

Now let us consider the case with Yukawa coupling to fermions. Unlike ordinary stellar objects, we are not restricted to the SM fermions. 
In the high-temperature or high-density environment within the interior of 2-2-holes, beyond the SM heavy fermions could exist due to either primordial production in the early universe and subsequent evolution, or production resulting from high-energy particle collisions in the interior. 
These fermions would then play a significant role in sourcing the scalar field. By taking the high temperature or density limit, specifically $T\gg m_f$ or $k_F \gg m_f$, we find that both effects are encoded by the $F$-term. From Eqs.~(\ref{eq:FGT}) and (\ref{eq:FGrho1}), we find the dimensionless counterparts as
\begin{eqnarray}
\bar{F}_T(\bar{r})&=&\frac{g_{\phi f}}{6} \frac{m_{f,0}}{\phi_0}\bar\lambda_2\bar{T}^2(\bar{r})
\equiv\mathcal{F}_T\, \bar{T}^2(\bar{r})\nonumber\\
\bar{F}_\rho(\bar{r})&=&\frac{g_{\phi f}}{2\pi^2} \frac{m_{f,0}}{\phi_0}\bar\lambda_2\bar{k}_F^2(\bar{r})
\equiv\mathcal{F}_\rho\, \bar{k}_F^2(\bar{r})\,,
\end{eqnarray}
where $\mathcal{F}_\rho=3\mathcal{F}_T/\pi^2= g_{\phi f} m_{f,0}\bar\lambda_2/(2\pi^2\phi_0)$ represent the dimensionless coefficients. 

\begin{figure}[!h]
	\centering
	\includegraphics[height=5.2cm]{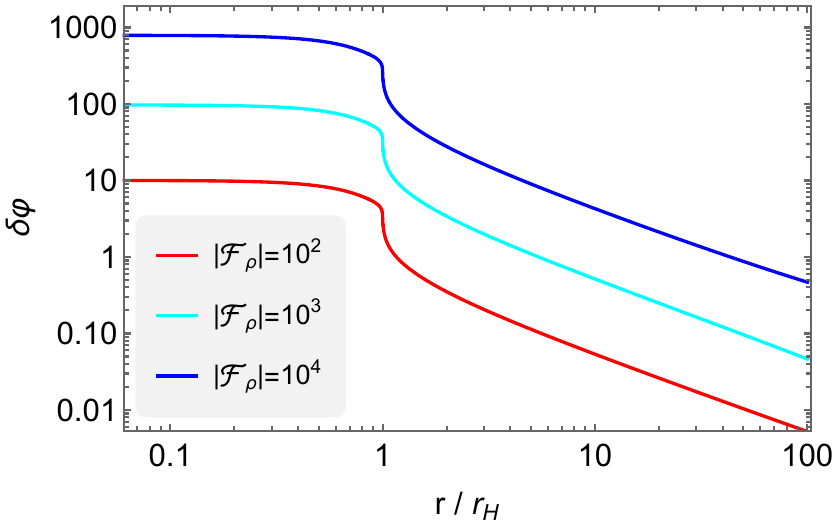}\;\;	
	\includegraphics[height=5.2cm]{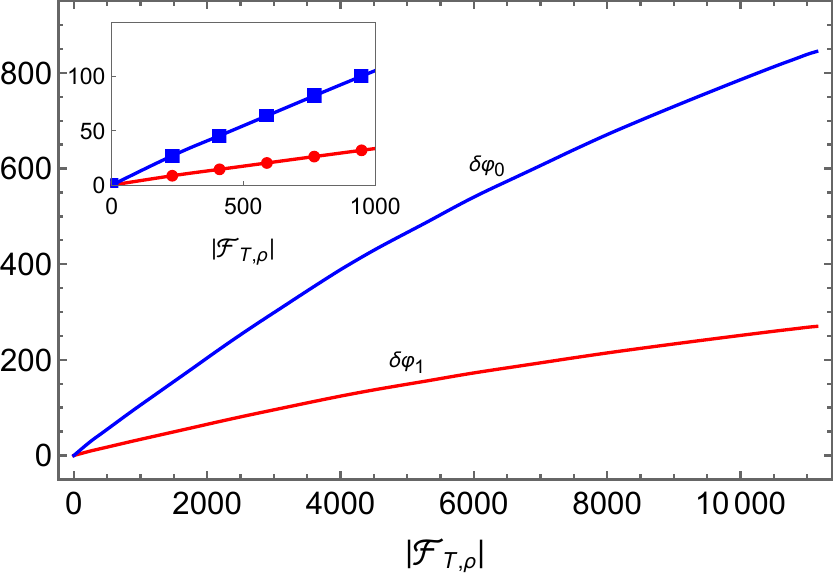}
\caption{\label{fig:22holeF} Finite temperature and density effects induced by the Yukawa couplings to fermions for 2-2-holes. Left: the rescaled scalar profile $\left|\delta\varphi\right|$ as a function of $\bar{r}\equiv r/r_H$  for several benchmark values of the dimensionless coefficient $|\mathcal{F}_{\rho}|$. Right: rescaled scalar field values $\delta\varphi_0$ and $\delta\varphi_1$ as functions of $|\mathcal{F}_{T,\rho}|$. }
\end{figure}

Figure~\ref{fig:22holeF} illustrates the numerical results for this case. The scalar field again experiences a more pronounced decrease at $\bar{r}\sim 1$ as in Fig.~\ref{fig:22holeG}, due to the special feature of 2-2-holes. The difference between cold Fermi gas and photon gas around $\bar{r}$ close to 1 has a small impact on the profile.
Similar to Fig.~\ref{fig:OSF}, we observe that $\delta\varphi_0$ and $\delta\varphi_1$ are linear in the magnitude $|\mathcal{F}_{T,\rho}|$ for small values and continue to increase with $|\mathcal{F}_{T,\rho}|$ for large values. 
The ratio $\delta\varphi_1/\delta\varphi_0$ is approximately half of that in Fig.~\ref{fig:OSF}, related to the notable decrease in the scalar profile at $\bar{r}\sim 1$.

\begin{figure}[!h]
    \centering
    \includegraphics[height=6cm]{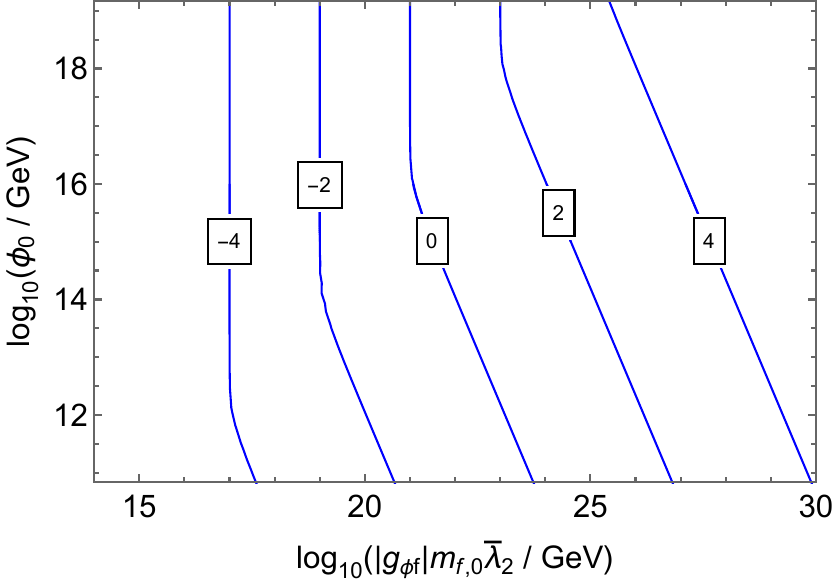}
    \caption{Contours of the scalar charge-to-mass ratio $\gamma$ on the plane of the scalar VEV $\phi_0$ and the combination $|g_{\phi f}|m_{f,0}\bar{\lambda}_2$.}
    \label{fig:contour22holeF}
\end{figure}

The contour of $\gamma$ in Fig.~\ref{fig:contour22holeF} shows a mild dependence on $\phi_0$ and becomes completely independent of $\phi_0$ in the limit of small $|\mathcal{F}_{T,\rho}|$, with $\gamma\approx 0.01 |g_{\phi f}|m_{f,0}\bar{\lambda}_2/\Mp$.
To achieve a value of $\gamma$ on the order of 1, the combination $|g_{\phi f}|m_{f,0}\bar{\lambda}_2$ needs to reach $10^{22}\,$GeV. Considering that $|g_{\phi f}|\lesssim 1$, this implies $\bar{\lambda}_2\gtrsim 10^{12}(10^{10}\textrm{GeV}/m_{f,0})$. 
Namely, for $m_{f,0}\sim 10^{13}\,$GeV and $|g_{\phi f}|\sim 10^{-3}$, a value of $\gamma\sim1$ can be achieved with $\bar{\lambda}_2\gtrsim 10^{12}$, corresponding to a minimum 2-2-hole mass of $\Mmin\gtrsim 10^7\,$g.\footnote{For reference, when considering $\bar\lambda_2\sim 10^{12}$ and $m_{f,0}\sim 10^{13}\,$GeV, we observe $m_f(\lambda_2\lp)^{1/2}\sim 1$, corresponding exactly to the benchmark value in Fig.~\ref{fig:22hole}.}

In comparison to the case with only self-interaction, the scalar mass is not directly involved in determining the scalar charge. Furthermore, the presence of a potentially large mass for the new heavy fermion and the absence of experimental constraints on $g_{\phi f}$ allow for more freedom in achieving a large scalar charge, which compensates for the small radial size of the 2-2-hole interior. Thus, astrophysical 2-2-holes ($M\gtrsim M_{\odot}$) with non-trivial scalar profiles could indeed have observational effects on ongoing and planned gravitational wave experiments.
It is important to note that the above estimate assumes that the heavy fermion constitutes all of the gas sourcing the 2-2-holes, while in reality the source should also include the SM fermions. Given the wide range of allowed parameter space shown in Fig.~\ref{fig:22holeF}, it is evident that even a small fraction of heavy fermions is sufficient to achieve a significant charge.

Therefore, the potential existence of 2-2-holes opens up a new possibility for generating non-trivial light scalar profiles in the strong gravity regime.
The deep gravitational potential within the interior of these holes leads to extremely high temperatures or densities of the gas, resulting in a significantly modified scalar potential. However, the high curvature effects in the interior cause a reduction in the effective radial size of the object, leading to an additional suppression of the scalar charge compared to cases in GR. By combining 
these two effects, we find that a magnitude of $\gamma\sim 1$ can be universally achieved for 2-2-holes of all masses $M\gg M_{\rm min}$ if the scalar field couples to some new heavy fermions, taking advantage of the exceptionally hot and dense environment provided by their interiors.

\section{Gravitational wave observations of scalarized 2-2-holes}
\label{sec:GWobs}

The possibility that astrophysical 2-2-holes could have a significant scalar charge in the minimal model raises an interesting question about the observational consequences of scalarized 2-2-holes.
As our exploration in this paper is limited to the test field limit, we are unable to study constraints arising from changes in the background spacetime due to the backreaction of the scalar profile. Nevertheless, we anticipate that the electromagnetic observations of Sgr\,A$^*$ from Event Horizon Telescope have the potential to exclude the presence of scalar charge with a charge-to-mass ratio $\gamma^2\lesssim \mathcal{O}(0.1)$, as indicated by studies in various scenarios in Ref.~\cite{Vagnozzi:2022moj}.
Thus, our subsequent discussion will focus on probing binary systems involving either one or two scalarized 2-2-holes with a smaller $\gamma$ using gravitational wave observations. Additionally, we only consider the inspiral stage, where the two objects are far apart, i.e. $r\gg r_H$. In this regime, the potential backreaction from the scalar field is also negligible.

Below, we will first discuss the inspiral dynamics of the binary system with scalar charges, and then briefly explore potential observations for specific cases in the scenario where all astrophysical black holes are 2-2-holes. For demonstration purposes, we consider observational effects for two types of systems: the stellar-mass binaries and extreme mass-ratio inspirals (EMRIs), which involve supermassive 2-2-holes at the center of galaxies.
It is important to note that the difference in the $M$ dependence of the scalar charge for 2-2-holes, compared to scalarized black holes in scalar-tensor theories, has significant implications for observations.

The presence of scalar charges within a binary system is anticipated to generate additional scalar forces or scalar radiations. For simplicity, we focus on the inspiral stage with a radius much larger than the innermost stable circular orbit (ISCO) at $r_{\rm ISCO}\approx 3r_H$, where the dynamics can be adequately described at the leading-order Newtonian approximation. 
In the center of mass frame, the dynamics can be described as a one-body problem with the total mass  $M=M_1+M_2$, the reduced mass $\mu=M_1M_2/M$, and the relative coordinate $\bf{r}=r_1-r_2$, where the reduced mass is related to the mass ratio $q=M_2/M_1$ by $\mu/M=q/(1+q)^2$.
Here, we focus on circular orbits, and the kinetic variables of interest are the orbital radius $r=|\bf{r}|$ and the orbital frequency $\Omega$. 
For relatively small scalar charge $\gamma\lesssim \mathcal{O}(0.1)$ and large distance $r\gtrsim r_{\rm ISCO}$, it suffices to consider quasi-circular orbits with slowly varying $r$, namely, when $\dot{\Omega}\ll\Omega^2$.    
The total energy of the system is then given by the sum of the kinetic energy and potential energy from gravity and scalar force, i.e. 
\begin{eqnarray}  \label{total  energy}
E_{\rm orb}=\frac{1}{2}\mu\, r^2\Omega ^2+V_{\rm orb},\quad
V_{\rm orb}=-\frac{\lp^2 M\mu}{r}\left( 1+\gamma_1\gamma_2 \mathrm{e}^{-m_{\phi}r} \right), 
\end{eqnarray}
The orbital frequency $\Omega$ is related to $r$ by the modified Kepler relation, which is given by $d V_{\rm orb}/dr=\mu\, r\, \Omega^2$, namely,
\begin{equation}
    \Omega ^2=\frac{\lp^2 M}{r^3}\left[ 1+\mathrm{e}^{-m_{\phi}r}\gamma_1\gamma_2 \left( 1+m_{\phi}r \right) \right]\,. 
\end{equation}
In the massless limit, i.e.  $m_\phi r\ll 1$, the modified Kepler relation becomes $\Omega^2\approx \lp^2 M(1+\gamma_1\gamma_2 )/r^3$, and the total energy deduces to the simple form $E_{\rm orb}\approx-\lp^2 M\mu\left(1+\gamma_1\gamma_2  \right)/(2r)$. It is then easy to verify that the assumption of quasi-circular orbits remains good for $\gamma_i\lesssim \mathcal{O}(0.1)$ and $r\gtrsim r_{\rm ISCO}$.\footnote{The condition $\dot{\Omega}\ll\Omega^2$ requires that $\frac{6\sqrt{2}}{5} (r/r_H)^{-5/2}( 1+\gamma_1\gamma_2 ) ^{3/2}( 1+\frac{1}{12}\gamma_1\gamma_2 ) \ll 1$. Considering that $r\gtrsim r_{\rm ISCO}=3r_H$, this implies that $\gamma_1\gamma_2\ll 2.8$. Thus, the assumption of quasi-circular orbits is valid for $\gamma_i\lesssim \mathcal{O}(0.1)$.}

Apart from gravitational waves radiation, the binary system of scalarized objects  can also emit scalar radiations when $\Omega\gtrsim m_\phi$. As the scalar radiation starts from dipole radiation, we consider the spherical harmonic expansion up to the quadrupole order, i.e. $\ell=2$. The orbital evolution can be then approximately determined by
\begin{eqnarray}
\frac{\mathrm{d} E_{\rm orb}}{\mathrm{d}t}=-P_{\rm GW}-P^{(\ell=1)}_{\rm SR}-P^{(\ell=2)}_{\rm SR}\,,
\end{eqnarray}
with the radiated powers~\cite{Huang:2018pbu,Luu_2024}
\begin{eqnarray}  \label{energy loss}
\begin{aligned}
P_{\rm GW}&=\frac{32}{5}\lp^2\mu ^2r^4\Omega ^6,
\\
P^{(\ell=1)}_{\rm SR}&=\frac{1}{12\pi}\frac{\left( Q_1M_2-Q_2M_1 \right) ^2}{M^2}r^2\Omega ^4\left( 1-\frac{m_{\phi}^{2}}{\Omega ^2} \right) ^{3/2},
\\
P^{(\ell=2)}_{\rm SR}&=\frac{4}{15\pi}\frac{\left( Q_1M_{2}^{2}+Q_2M_{1}^{2} \right) ^2}{M^4}r^4\Omega ^6\left( 1-\frac{m_{\phi}^{2}}{4\Omega ^2} \right) ^{5/2}.
\end{aligned}
\end{eqnarray}

To further simplify the evolution equations, it is useful to consider two concrete cases:
\begin{itemize}
    \item \textbf{Case A}: a binary consisting of two scalarized 2-2-holes with approximately equal scalar charge-to-mass ratios, i.e. $\gamma_1=\gamma_2\equiv \gamma$
    \item \textbf{Case B}: a binary consisting of one scalarized 2-2-hole with nonzero $\gamma$ and one ordinary stellar object with negligible scalar charge
\end{itemize}
For case A, the scalar dipole moment vanishes, and rescaled evolution equations are simplified as 
\begin{eqnarray}\label{eq:Evolution1}
\bar{E}_{\rm orb}=\frac{1}{2} \bar{r}^2\bar{\Omega} ^2-\frac{1}{2\bar{r}}\left( 1+\gamma^2 \mathrm{e}^{-\eta \bar{r}} \right),
\quad
\frac{\mathrm{d} \bar{E}_{\rm orb}}{\mathrm{d}\bar{t}}=-\frac{16}{5}\bar{r}^4\bar\Omega ^6-\frac{8}{15}\gamma^2 \bar{r}^4\bar\Omega ^6\left( 1-\frac{\eta^{2}}{4\bar\Omega ^2} \right)^{\frac{5}{2}},
\end{eqnarray}
where $\bar{r}=r/r_{\mathrm{H}}$, $\bar\Omega=\Omega r_{\mathrm{H}}$, and $\bar{E}_{\rm orb}=E_{\rm orb}/\mu$, $\bar{t}=(t/r_{\mathrm{H}})(\mu/M)$. 
Note that $\bar{t}$ is defined slightly differently here in order to incorporate the $\mu$ or $q$ dependence. This illustrates that variations in $q$ only affect the rate of evolution. 
The modified Kepler relation becomes: $\bar\Omega^2=(1+e^{-\eta \bar{r}}\gamma^2(1+\eta\bar{r}))/(2\bar{r}^3)$. In such systems,  the scalar charge generates an additional Yukawa force as well as an additional quadruple radiation. These effects are both proportional to $\gamma^2$.  
For case B, there is no additional scalar force, and  the leading order corrections come from the scalar dipole radiation. The rescaled evolution equations then become,
\begin{eqnarray}\label{eq:Evolution2}
\bar{E}_{\rm orb}=\frac{1}{2} \bar{r}^2\bar{\Omega} ^2-\frac{1}{2\bar{r}},
\quad
\frac{\mathrm{d} \bar{E}_{\rm orb}}{\mathrm{d}\bar{t}}=-\frac{16}{5}\bar{r}^4\bar\Omega ^6-\frac{1}{6}\gamma^2 \bar{r}^2\bar\Omega ^4\left( 1-\frac{\eta^{2}}{\bar\Omega ^2} \right)^{\frac{3}{2}},
\end{eqnarray}
with the standard Kepler relation  $\bar\Omega^2=1/(2\bar{r}^3)$.

To solve for the orbital evolution, we first determine $\bar\Omega(\bar{r})$ using the modified Kepler relation. We then solve for $\bar{r}(\bar{t})$ by substituting $\bar\Omega(\bar{r})$ into Eqs.~(\ref{eq:Evolution1}) and (\ref{eq:Evolution2}). In general, the scalar corrections are considered negligible at sufficiently large distances, where $\eta \bar{r}\gg 1$ or $\eta\gg \bar\Omega$, and the evolution can be described approximately by GR. On the other hand, at sufficiently close distances, the zero mass limit applies. To retain the explicit dependence on $\eta$, the orbital evolution must be solved numerically.

Analytical solutions can be obtained in the zero mass limit, i.e. $\eta\bar{r}\ll 1$.
For case A, we find 
$d\bar{\Omega}/d\bar{t}\approx 6\left( 1+\gamma ^2 \right) ^{\frac{2}{3}}( 1+\frac{1}{6}\gamma ^2) \bar{\Omega}^{\frac{11}{3}}$ 
from Eq.~(\ref{eq:Evolution1}), where the factors  $1+\gamma^2$ and $1+\frac{1}{6}\gamma ^2$ represent the corrections from the scalar force and scalar radiation, respectively. 
Directly integrating out this expression from $\bar{t}=0$, we obtain 
\begin{equation} \label{eq:rtbarA}
  \bar{\Omega}\left( \bar{t} \right) \approx  \left[ \bar{\Omega}_{0}^{-\frac{8}{3}}-16\left( 1+\gamma ^2 \right) ^{\frac{2}{3}}\left( 1+\frac{\gamma ^2}{6} \right) \bar{t} \right] ^{-3/8},
\end{equation}
where $\bar{\Omega}_0=\bar{\Omega}(\bar{t}=0)$. 
For case B, we find $d\bar{\Omega}/d\bar{t}\approx 6 \bar{\Omega}^{\frac{11}{3}}+\frac{1}{2}\gamma^2\bar\Omega^3$
from Eq.~(\ref{eq:Evolution2}), where the $\bar\Omega^3$ term represents the dipole contribution. Because of its different $\bar{\Omega}$ dependence, $\bar{\Omega}(\bar{t})$ can only be solved implicitly. Below, we present the result in terms of a small $\gamma$ expansion,
\begin{eqnarray}\label{eq:rtbarB}
\bar{t}\left( \bar{\Omega} \right) 
\approx \frac{1}{16}\left( \bar{\Omega}_0^{-\frac{8}{3}}-\bar{\Omega}^{-\frac{8}{3}} \right)-\frac{1}{240}\gamma ^2\left( \bar{\Omega}_0^{-\frac{10}{3}}-\bar{\Omega}^{-\frac{10}{3}} \right) +\mathcal{O} \left( \gamma ^4 \right)\,.
\end{eqnarray}
In general, starting from some initial value, the frequency gradually increases with time and then quickly increases as it approaches the ISCO. The evolution becomes faster with larger $\gamma^2$.


Next, let us consider observations for specific systems. 
First, we focus on stellar-mass binaries with a total mass $M$ ranging from $\mathcal{O}(10M_{\odot})$ to $\mathcal{O}(100M_{\odot})$. 
Joint observations of these binaries using both ground-based and space-based detectors have been proposed as a powerful way to improve the constraints on the scalar charges of black holes~\cite{Barausse:2016eii}. The key observable for such estimations is the coalescence time $t_{\rm coal}$, which represents the time it takes for the binary to merge within the LIGO band from an initial $\Omega_0$ in the 
millihertz band of space-based detectors. 
Assuming that the binary can be observed with exceptional precision using the latter, the merger time in the LIGO band can be accurately predicted. This prediction can then be utilized to constrain modifications of $t_{\rm coal}$
due to the scalar charges.

\begin{figure}[H]
    \begin{minipage}{0.5\textwidth}
      \centering
      \includegraphics[height=5.2cm]{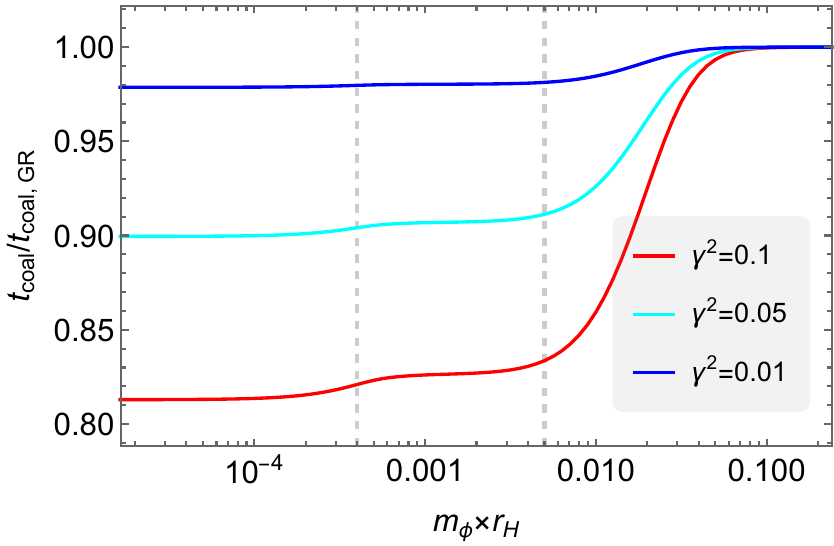}
      \end{minipage}
       \begin{minipage}{0.5\textwidth}
      \centering
      \includegraphics[height=5.2cm]{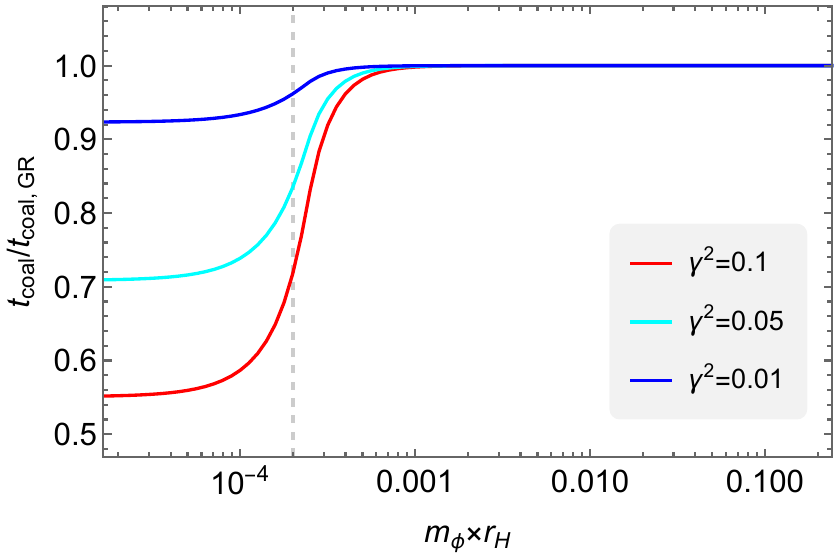}
      \end{minipage}
    \caption{The coalescence time ratio as a function of $\eta=m_\phi r_H$ for different values of scalar charge-to-mass ratio $\gamma^2$. Left: case A, where the two dashed vertical lines denote $2\bar\Omega_0$ and $1/\bar{r}_0$ from left to right. Right:  case B, where the dashed vertical lines denote $\bar\Omega_0$. In both panels, the initial frequency is set as $\bar{\Omega}_0=2\times 10^{-4}$.}
    \label{fig:t_coal}
\end{figure}

For systems involving scalarized 2-2-holes, in the massless limit, the rescaled coalescence time $\bar{t}_{\rm coal}$ can be estimated for the two cases by setting $\bar{r}(\bar{t}_{\rm coal})\approx 0$ in Eqs.~(\ref{eq:rtbarA}) and (\ref{eq:rtbarB}). This yields 
\begin{eqnarray}\label{eq:tcoal1}
 \bar{t}_{\mathrm{coal}}\approx\bar{t}_{\mathrm{coal, GR}}\left\{\begin{array}{ll}
 \left( 1+\gamma ^2 \right) ^{-\frac{2}{3}}\left( 1+\frac{1}{6}\gamma ^2\right)^{-1}\,, & \textrm{Case A}\\
 1-\frac{1}{15}\gamma ^2\bar{\Omega}_0^{-\frac{2}{3}}\,, & \textrm{Case B}
  \end{array}\right.\,,
\end{eqnarray}
where $\bar{t}_{\mathrm{coal, GR}}=\frac{1}{16}\bar{\Omega}_{0}^{-8/3}$
represents the prediction in GR. 
In contrast to case A, the corrections for case B are amplified by a factor of $\bar{\Omega}_0^{-2/3}\gg1$, due to the dominance of dipole radiation during the early inspiral stage. 
Fig.~\ref{fig:t_coal} shows the coalescence time ratio as a function of scalar mass parameter $\eta$. For case A, the results tend towards the massless limit described in Eq.~(\ref{eq:tcoal1}) when $\eta\ll 2\bar\Omega$. Conversely, they approach the GR limit when $\eta\gg 1/\bar{r}_0$, where both the force and radiation effects are suppressed. In the intermediate region, the scalar force is effective from the beginning, while the radiation only becomes significant at a later time. Comparing the two effects, the scalar force has a larger contribution. Similarly, in case B, we observe a convergence towards the massless limit outlined in Eq.~(\ref{eq:tcoal1}) and the GR limit when $\eta\ll \bar\Omega$ and $\eta\gg \bar\Omega$, respectively. As anticipated, for $\gamma^2\lesssim \mathcal{O}(0.1)$, we observe more pronounced changes in case B compared to case A.

To provide a rough estimate, let us consider a GW150914-like binary consisting of scalarized 2-2-holes.
This binary takes about five years to evolve from the LISA band with a frequency of $\sim 0.01\,$Hz to LIGO band at $\sim 100\,$Hz. Assuming that $t_{\mathrm{coal}}$ can be predicted up to 10\,sec through observations from LISA~\cite{Barausse:2016eii}, the constraint on the scalar charge-to-mass ratio can be obtained by imposing the condition $t_{\mathrm{coal}}-t_{\mathrm{coal},\mathrm{GR}}\lesssim 10\,$sec. The most stringent limits are expected for $m_\phi\lesssim 10^{-15}\,$eV from Fig.~\ref{fig:t_coal}, where the scalar is effectively massless during the evolution from LISA to LIGO. Utilizing the expressions for $t_{\mathrm{coal}}$ in Eq.~(\ref{eq:tcoal1}) for the massless case, this condition yields a limit of $\gamma^2\lesssim 10^{-8}$ for case A and $\gamma^2\lesssim 10^{-9}$ for case B, respectively. The limit for case B is about 1 order better due to the enhanced effect of scalar dipole radiation during the earlier inspiral stage, where $\bar{\Omega}_0^{-2/3}\gg1$. 
However, it is important to consider the presence of degeneracies among the waveform parameters. Ref.~\cite{Barausse:2016eii} demonstrates that the limit on $\gamma^2$ could be 1 order worse for the case of dipole radiation due to these degeneracies.
Additionally, recent findings indicate that the prediction of $t_{\mathrm{coal}}$ may not be as accurate, with an uncertainty of $\sim 3\,$hours~\cite{Klein:2022rbf}. This would further deteriorate the limit on $\gamma^2$ by 3 orders of magnitude. A conservative estimate for the limit in case B would then be $\gamma^2\lesssim 10^{-5}$. In case A, the degeneracy among waveform parameters could be stronger, resulting in a significantly worse bound for $\gamma^2$. However, considering the possibility of a mild dependence of $\gamma^2$ on the mass $M$ due to the contribution from the transition region of 2-2-holes, the cancellation of dipole radiation may not be exact in case A. This can potentially aid in breaking the degeneracy. 
Furthermore, the possibility of enhancing the sensitivity by incorporating space-based detectors in the decihertz band has also been explored~\cite{Liu:2020nwz}.

As the second example, we consider EMRIs containing a supermassive scalarized 2-2-hole  with mass $\sim (10^4-10^7)M_\odot$ and a stellar mass object. The latter could either be an ordinary stellar object with negligible charge or a scalarized 2-2-hole.
Because of the small mass ratio $q$, the system undergoes a slow evolution in the millihertz band of space-based detectors over a span of months to years. This allows for the accumulation of small phase differences, making it possible to detect small deviations from GR.
Specifically, under the adiabatic approximation,
the gravitational wave phase of the dominant mode during the inspiral stages, which is twice the orbital phase, can be expressed approximately as
\begin{eqnarray}\label{eq:phiGW}
  \phi_{\rm GW}(t)=\int_{0}^{t} 2\Omega(t')dt'
  \approx \frac{2}{q}\int_{\bar{\Omega}_0}^{\bar{\Omega}(t)} \bar\Omega'\left(\frac{d\bar{\Omega}'}{d\bar{t}'} \right) ^{-1}d\bar{\Omega}'.
\end{eqnarray}
A useful measure for estimating the effects of a scalar field is the accumulated dephasing between the cases with and without a scalar profile. 
This dephasing is determined as the difference in the gravitational wave phase for a given $\Omega_0$ and $t$, i.e. 
\begin{eqnarray}\label{eq:dephasing}
  \Delta\phi_{\rm GW}(t)=\phi_{\rm GW, GR}(t)-\phi_{\rm GW}(t)\,,
\end{eqnarray}
where $\phi_{\rm GW, GR}(t)=\frac{1}{5q}\big[ \bar{\Omega}_{0}^{-5/3}-\bar{\Omega}(t)^{-5/3}\big]$ denotes the GR prediction.

\begin{figure}[!h]
      \centering
      \includegraphics[height=4.9cm]{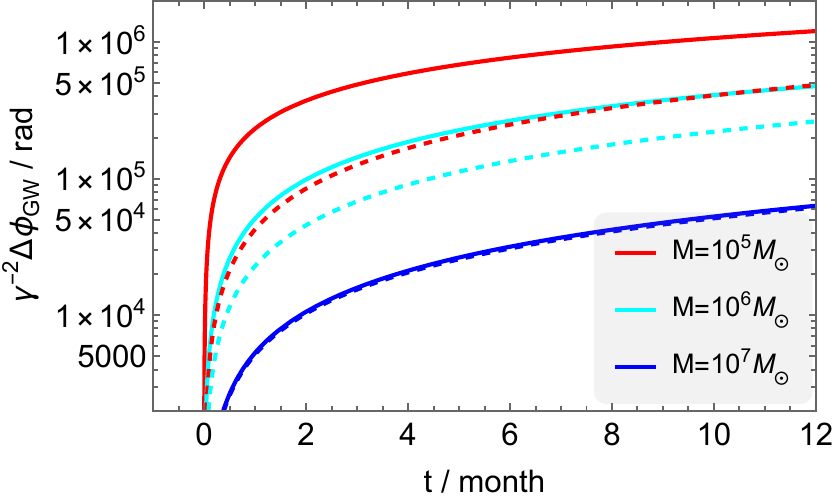}\;\;
      \includegraphics[height=4.9cm]{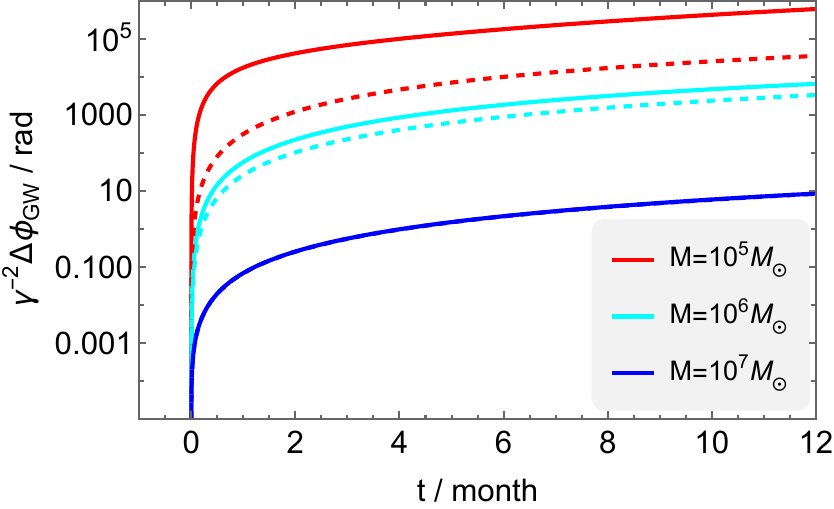}\;\;
    \caption{The normalized dephasing $\gamma^{-2}\Delta\phi$ for an EMRI system with different total mass $M$ and coalescence time $t_{\rm coal}$.  Left: case A, with the secondary mass $M_2\approx 10M_\odot$. Right: case B, with the secondary mass $M_2\approx 1.4M_\odot$ as for a typical NS. In both panels, the solid and dashed lines represent $t_{\rm coal}\approx 1$\,yr and $t_{\rm coal}\approx 4$\,yr, respectively.}
    \label{fig:DeltaN}
\end{figure}

In contrast to stellar mass-binaries, EMRIs evolve extremely slow in the observational band. Hence, for the majority of parameter space, we are either in the regime that the suppression due to the mass is too strong or the regime that the massless limit is good. To derive the most stringent limits, we focus on the massless limit. By substituting Eqs.~\eqref{eq:Evolution1} and  \eqref{eq:Evolution2} into Eqs.~(\ref{eq:phiGW}) and (\ref{eq:dephasing}), 
we can obtain the dephasing for the two cases under the small $\gamma^2$ expansion
\begin{eqnarray}  \label{eq: dephasing1}
  \Delta\phi_{\rm GW}(t)\approx \left\{\begin{array}{ll}
 \frac{1}{10}\frac{\gamma^2}{q}\big[ \bar{\Omega}_{0}^{-5/3}-\bar{\Omega}(t)^{-5/3}\big]\,, & \textrm{Case A}\\
\frac{1}{84}\frac{\gamma^2}{q}\big[ \bar{\Omega}_{0}^{-7/3}-\bar{\Omega}(t)^{-7/3}\big]\,, & \textrm{Case B}
  \end{array}\right.\,. 
\end{eqnarray}
Similarly, there is an additional factor of $\bar{\Omega}_0^{-2/3}$ for case B compared to case A. 
As the dephasing is proportional to $\gamma^2$ in the small charge limit, we display in  Fig.~\ref{fig:DeltaN} the numerical results of $\gamma^{-2}\Delta\phi_{\rm GW}(t)$ for EMRIs of different mass $M$ and $t_{\rm coal}$. For both cases, the dephasing grows more rapidly with time for the smaller $M$ case, as a larger cycle number has been accumulated within the given timescale. Case B shows a much stronger dependence on $M$ due to the additional $\bar{\Omega}_0^{-2/3}$ factor. In terms of the $t_{\rm coal}$ dependence, the dephasing at a given $t$ is reduced for the EMRI with a longer $t_{\rm coal}$,or a smaller $\Omega_0$. 
This aligns with the leading order expansion of Eq.~(\ref{eq: dephasing1}) in the small $\bar{t}$ limit, where $\Delta\phi_{\rm GW}(t)\propto \bar{t}_{\mathrm{coal}}^{-3/8}\bar{t}$ for case A and $\Delta\phi_{\rm GW}(t)\propto \bar{t}_{\mathrm{coal}}^{-1/8}\bar{t}$ for case B. 
The contrast between the two coalescence time cases is more pronounced for EMRIs with a smaller $M$.

As for a rough estimate, let us consider an EMRI with a supermassive scalarized 2-2-hole of $M\sim 10^{6}M_{\odot}$. Assuming an average signal-to-noise ratio (SNR) of detected events of approximately $30$, a dephasing of $\Delta \phi_{\rm GW}\sim 0.1\,$rad is considered to be detectable~\cite{Lindblom:2008cm, Bonga:2019ycj}. For an EMRI with $t_{\rm coal}\approx 1$\,yr, corresponding to an accumulation time of one year, 
the scalar charge would be constrained to $\gamma^2\lesssim 10^{-7}$ for case A and $\gamma^2\lesssim 10^{-5}$ for case B from Fig.~\ref{fig:DeltaN}. Additionally, the rapid spin of the supermassive 2-2-hole, caused by accretion, is expected to significantly increase the orbital frequency overall, leading to a reduction in the dephasing as predicted in Eq.~(\ref{eq: dephasing1}). 
This results in worse constraints compared to the non-spinning case, i.e. $\gamma^2\lesssim 6\times 10^{-7}$ for case A and $\gamma^2\lesssim 10^{-4}$ for case B. 
Finally, the sensitivity will be significantly compromised due to degeneracies with other waveform parameters. For case B with $\chi\sim 0.9$ and $M\sim 10^6M_{\odot}$, a Fisher analysis has demonstrated that only $\gamma^2\lesssim 0.03$ is expected at the $1\sigma$ level after one year of observation on LISA with $\mathrm{SNR}\sim 150$~\cite{Maselli:2021men}.

\section{Summary}
\label{sec:summary}

In this paper, we investigate a novel method for generating long-range scalar forces that exclusively manifest around astrophysical black holes. 
If all observed black holes are horizonless and ultracompact 2-2-holes, which are potential end points of gravitational collapse in quadratic gravity, the hot or dense gases inside these UCOs allow for the generation of non-trivial scalar profiles with a significant charge through environmental effects. This is in contrast to other scenarios, where either non-minimal coupling to gravity or violation of energy conditions are required.

For demonstration purposes, in this work, we focus on a minimal model of the scalar field with a double-well scalar potential and Yukawa interaction with fermions. In Sec.~\ref{sec:scalar}, we investigate the effects of finite temperature and density on the scalar potential. We find that these corrections can be effectively described by either a linear term or a quadratic term, as shown in Eq.~(\ref{eq:fullV2}). This leads to two mechanisms for generating non-trivial scalar profiles, as demonstrated in Fig.~\ref{fig:VphiFG}.
In Sec.~\ref{sec:stellar}, we further examine the non-trivial scalar profile and the predicted scalar charge in the test field limit, for the minimal model in both ordinary stellar objects and 2-2-holes. We observe that the scalar charge of typical ordinary stellar objects is significantly suppressed in the minimal model, making it difficult to probe using even the high precision fifth-force measurements.
In contrast, 2-2-holes have the unique capability of sourcing light scalar fields. The exceptionally high temperatures or densities of the gases within their interior allow for the generation of a significant scalar charge through environmental effects. As depicted in Fig.~\ref{fig:22holeF}, we demonstrate that a scalar-to-charge mass ratio $\gamma$ of order 1 can be readily achieved if a considerable fraction of new heavy fermions within the 2-2-hole interior couple to the scalar field.

In Sec.~\ref{sec:GWobs}, we investigate the gravitational wave observations of scalarized 2-2-holes in the test field limit. The unique scaling of 2-2-holes results in their scalar charge scaling linearly with their mass, yielding a nearly constant value of $\gamma$ across a wide range of masses. This is in contrast to scalarized black holes in scalar-tensor theories, where smaller black holes exhibit significantly larger charges~\cite{Maselli:2020zgv}. Consequently, a binary system consisting of two 2-2-holes  experiences an additional scalar force and emits additional quadrupole scalar radiation, while a binary involving one 2-2-hole and one ordinary stellar object is primarily influenced by the dipole radiation.
For the former case, the value of $\gamma$ can be effectively constrained through multi-band gravitational wave observations of stellar-mass binaries of 2-2-holes. In the latter case, $\gamma$ can be probed through precise observations of EMRIs involving a supermassive 2-2-hole with space-based detectors.

In this work, we investigate the non-trivial scalar profile of 2-2-holes in the test field limit. As discussed in more detail in the Appendix~\ref{app:backreaction}, the backreaction of the scalar field might introduce more considerable effects in the vacuum regime. Therefore, it would be intriguing to explore the fully non-linear solution of scalarized 2-2-holes.
Additionally, our current study has not accounted for the contribution of the transition region around the would-be horizon of 2-2-holes.
Further research is warranted to explore the role of this region, especially in the context of the non-linear solution, where an interplay between the high curvature terms and scalar charge would be anticipated. 
Finally, the minimal model predicts a positive scalar charge for 2-2-holes, whereas adopting a more complex scalar potential could result in 2-2-holes with opposite charges. This possibility could lead to a wider range of observational implications for gravitational wave observations and warrants further investigation.

\vspace{0.1cm}
\section*{Acknowledgements} 
\vspace{-0.1cm}
We would like to thank Ufuk Aydemir for the early collaboration and valuable discussions. Additionally, we thank the anonymous referee for their insightful comments. X. Li and J. Ren are supported in part by the National Natural Science Foundation of China under Grant No. 12275276.


\appendix


\section{Backreaction of scalar field and no-scalar-hair theorems}
\label{app:backreaction}

In the main text, we have ignored the backreaction of the scalar field on the metric. This approximation is valid for ordinary stars, where the scalar charge is significantly suppressed and the weak gravity expansion is applicable just outside of the stars. However, this assumption needs to be carefully evaluated for UCOs such as 2-2-holes. On one hand, they may possess a substantial charge, potentially resulting in a non-negligible scalar charge-to-mass ratio $\gamma$ of approximately 1. On the other hand, it is known that the test field approximation for the scalar field breaks down near the horizon, regardless of how small the scalar charge is~\cite{Herdeiro:2015waa}. In the Appendix, we will examine this approximation for 2-2-holes, which closely resemble black holes just outside the would-be horizon at a small distance. We will also discuss how the no-scalar-field theory is circumvented in this context.

For a static and spherically symmetric spacetime, with the line element in Eq.~(\ref{metrics}), the proper energy density and pressure are given by $\rho=-T_t^t=-T_{tt}/B$ and $P=T^r_r=T_{rr}/A$, respectively. 
According to Noether's theorem, the energy-momentum tensor of the scalar field is given by
\begin{eqnarray} \label{energy-momentum tensor}
\begin{aligned}
T_{\phi, \mu \nu}=-\frac{2}{\sqrt{-g}}\frac{\delta (\sqrt{-g}\mathcal{L} )}{\delta g^{\mu \nu}}=\partial _{\mu}\phi \partial _{\nu}\phi -g_{\mu \nu}\left[ \frac{1}{2}g^{\alpha \beta}\partial _{\alpha}\phi \partial _{\beta}\phi +V\left( \phi \right) \right]\,,
\end{aligned}
\end{eqnarray}
where $\mathcal{L}=\mathcal{L}_{\phi}+\mathcal{L}_{f}$. The proper energy density and pressure for the scalar field are then expressed as 
\begin{eqnarray} \label{scalar rho and p}
\rho_\phi&=&\frac{1}{2A(r)}\left( \frac{\partial \phi}{\partial r} \right) ^2+ V\left( \phi \right) 
=\left[\frac{1}{2\bar{A}\left( \bar{r} \right)}\left( \frac{\partial \varphi}{\partial \bar{r}} \right)^2
+ \bar{V}\left( \varphi \right)\right]  \frac{\phi^2 _0}{\lambda^2_2}\nonumber\\
P_\phi&=&\frac{1}{2A(r)}\left( \frac{\partial \phi}{\partial r} \right) ^2- V\left( \phi \right) 
=\left[\frac{1}{2\bar{A}\left( \bar{r} \right)}\left( \frac{\partial \varphi}{\partial \bar{r}} \right)^2
- \bar{V}\left( \varphi \right)\right]  \frac{\phi^2 _0}{\lambda^2_2}\,.
\end{eqnarray}
In the last equality, we express the density and pressure as functions of the rescaled quantities, where the rescaled scalar potential $\bar{V}(\varphi)=\bar{F}\varphi+\frac{1}{4}(2\bar{G}-\zeta^2)\varphi^2+\frac{1}{8}\zeta^2\varphi^4$, with $\bar{F}$, $\bar{G}$ and $\zeta$ defined below the EOM Eq.~(\ref{eq:rescaleEOM22hole}). This naturally defines a rescaled proper energy density and pressure as
\begin{eqnarray}\label{eq:rescalerhoP}
  \bar\rho_\phi=\rho_\phi\lambda^2_2/\phi_0^2,\quad
  \bar{P}_\phi=P_\phi\lambda^2_2/\phi_0^2\,.
\end{eqnarray}

To investigate the backreaction of the scalar field, let us first consider the interior of the 2-2-hole, where the approximation can be easily justified by comparing the stress tensor of the scalar field and the matter source. Given the similarity of the proper temperature of the photon gas and the proper Fermi momentum for the cold Fermi gas deep inside 2-2-holes in Fig.~\ref{fig:22hole}, we use the cold Fermi gas for demonstration here. At the leading order of high density expansion, its proper energy density is given by
\begin{equation}
   \rho_{\rm gas}=
   \frac{\pi ^2}{30}\mathcal{N} k_{F}^{4}\sim \mathcal{O} \left( 10 \right) \bar{k}_{F}^{4}\frac{\Mp^{2}}{\lambda _{2}^{2}}\,,
\end{equation}
where $\mathcal{N}$ denotes the number of particle species. In the case of a non-negligible Yukawa coupling between the scalar and the Fermi gas, a significant scalar charge of 2-2-holes can be achieved. Considering the massless case for simplicity, the scalar potential in the deep interior, i.e. at $r\ll r_F$ where $r_F$ denotes the radius at which $k_F$ drops to zeros as shown in Fig.~\ref{fig:22hole}, is given approximately by $\bar{V}(\varphi)\approx \mathcal{F}_\rho \bar{k}_F^2(\bar{r})\varphi$. The ratio between the proper energy density of the scalar field and the Fermi gas can then be expressed as:
\begin{equation}  \label{eq:rho phi}
    \frac{\rho _{\phi}}{\rho _{\rm gas}}\sim \mathcal{O} \left( 0.1 \right) \times \frac{\phi^2 _0}{\Mp^2}\frac{\left( \partial _{\bar{r}}\varphi \right) ^2+2\mathcal{F}_\rho \bar{A}\left( \bar{r} \right)  \bar{k}_F^2(\bar{r})\varphi }{2\bar{A}\left( \bar{r} \right) \bar{k}_{F}^{4}}\,.
\end{equation}
Around the origin, $\varphi$ approaches a constant, and we have $\bar{A}\propto \bar{r}^2$ and $\bar{k}_F\propto 1/\bar{r}$. This leads to the inequality $\partial_{\bar{r}}\varphi\ll \bar{A}(\bar{r})\bar{k}_F^2(\bar{r})\varphi\ll \bar{A}(\bar{r})\bar{k}_F^4(\bar{r})$, indicating that the proper energy density of the scalar field is dominated by the potential energy due to environmental effects and remains much smaller than that of the Fermi gas. As $r$ increases, the gas density $\rho_{\rm gas}$ quickly declines and drops significantly at $r_F$. The kinetic energy of the scalar then starts to dominate and quickly surpasses $\rho_{\rm gas}$.

\begin{figure}[H]
\centering
      \includegraphics[height=6cm]{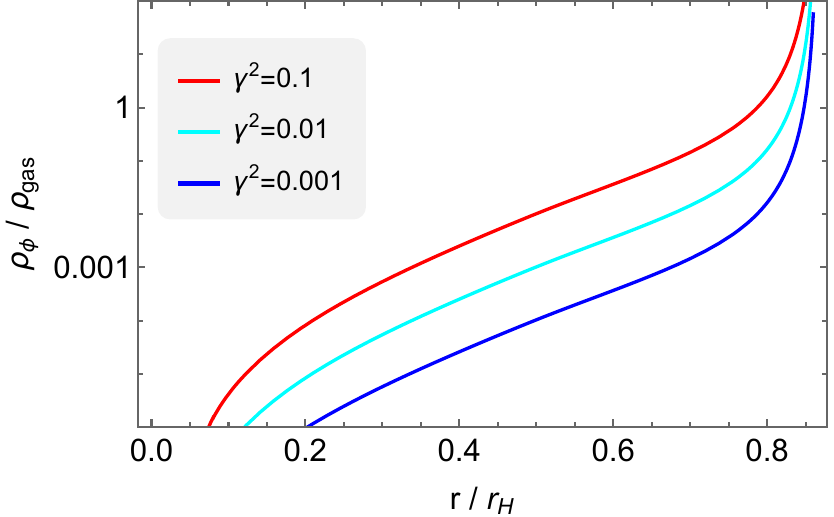}
    \caption{The proper energy density ratio of the scalar field and the cold Fermi gas in the 2-2-hole interior at different $\gamma^2$. Here, we choose $\phi_0=0.1m_{\rm pl}$ for demonstration.  For the $\gamma^2=0.1,0.01$ and $0.001$ cases, the dimensionless coefficient $|\mathcal{F}_\rho|\approx 106, 23, 11$, respectively. } 
    \label{fig:backreaction}
\end{figure}

Figure~\ref{fig:backreaction} displays the proper energy density ratio as a function of $\bar{r}$. For $\gamma^2=0.1$ and $\phi_0=0.1\Mp$, the contribution of the scalar field is significantly smaller than that of the Fermi gas at $r< r_F$, and therefore its backreaction can be safely ignored. It is important to note that for the chosen benchmark values of $\gamma^2$ and $\phi_0$, the magnitude of $|\mathcal{F}_\rho|$ remains small, where $\delta\varphi_1$ has a linear dependence as shown in Fig.~\ref{fig:22holeF}, and then the scalar charge $\gamma^2$ is approximately independent from $\phi_0$ in this regime.  If we were to consider a larger value of $\phi_0$, the corresponding $|\mathcal{F}_\rho|$ would be smaller, while $\gamma^2$ remains unchanged. Consequently, the ratio $\rho_\phi/\rho_{\rm gas}$ would scale as $(\phi_0/\Mp)^2$, as given by Eq.~(\ref{eq:rho phi}). To ensure the scalar field contribution is negligible in most of the region within $r<r_F$, it is then safe to consider $\phi_0$ not much larger than $0.1\Mp$.

Next, we will discuss the backreaction of the scalar field to the vacuum solution, where the contribution from the matter sources is negligible. This is applicable to the regime at $r\gtrsim r_F$, where $r_F$ is approximately $\mathcal{O}(0.1)r_H$, depending on the specific value of the gas mass. In the 2-2-hole exterior ($r\gtrsim r_H$) where the metric is well approximated by the Schwarzschild metric, the solution of scalar EOM in Eq.~(\ref{eq:EOMexterior}) in the massless limit is given by $\varphi(\bar{r})\propto \ln(1-1/\bar{r})$, which yields diverging $\rho_\phi$ and $P_\phi$  when $\bar{r}$ approaches 1. This indicates the breakdown of the test field approximation near the horizon for GR black holes~\cite{Herdeiro:2015waa}. 
For 2-2-holes, we would like to argue that this may not pose a significant issue, despite the fact that we have not yet obtained the full non-linear solution with the scalar profile due to numerical challenges.

Based on the numerical solutions of 2-2-holes without scalar charge, we observe that the quadratic curvature terms begin to dominate over the Einstein term just outside the would-be horizon. On the other hand, a full non-linear solution to the Einstein equations for a massless scalar field, which was established long ago by Fisher~\cite{Fisher:1948yn} and independently by Janis, Newman, and Winicour~\cite{Janis:1968zz}, exhibits a genuine curvature singularity at the modified horizon when the scalar charge is nonzero. The modified horizon radius depends on the scalar charge and is always greater than the Schwarzschild radius. When considering scalarized 2-2-holes with backreaction, we anticipate that the exterior will be described by the aforementioned non-linear solution outside of its modified horizon. Subsequently, the high curvature terms would take over and produce a different solution for the interior, similar to the zero charge case. As a result, the original curvature singularity at the horizon will be replaced by a high curvature interior, which approaches a timelike singularity at the origin. The latter has been argued to be a benign timelike singularity in the zero scalar charge case~\cite{Holdom:2016nek}.

\begin{figure}[!h]
      \centering
      \includegraphics[height=5.2cm]{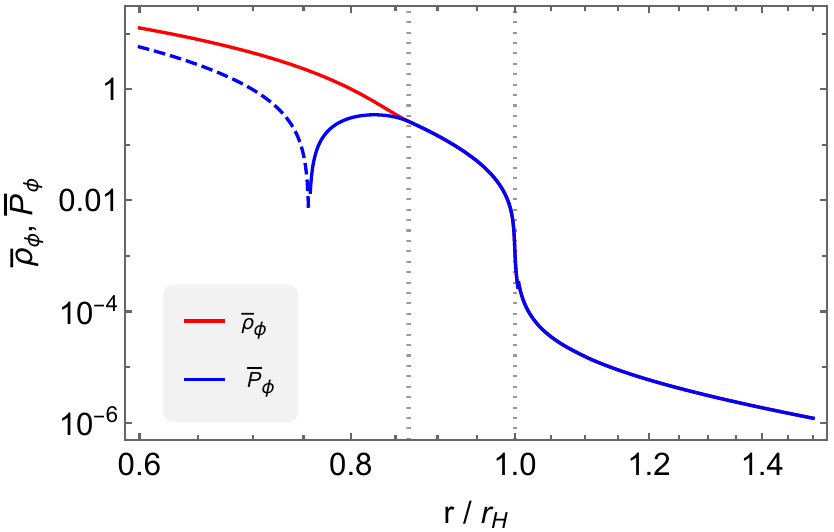}\;\;
      \includegraphics[height=5.2cm]{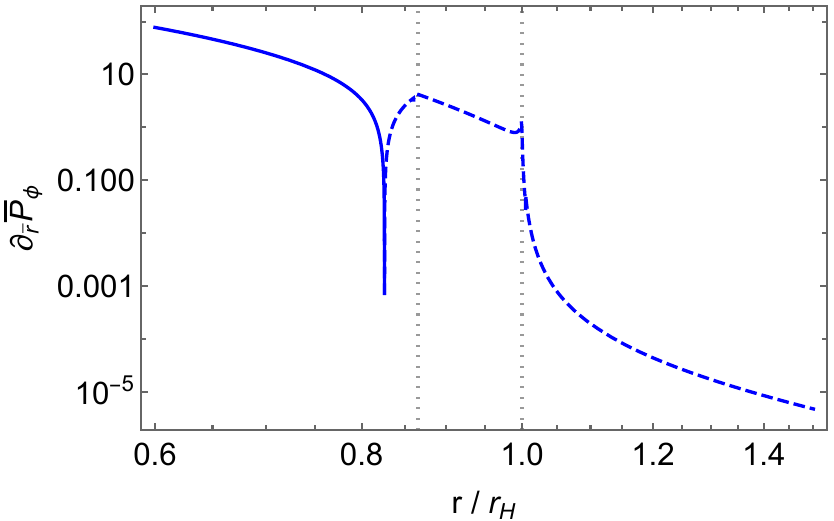}
    \caption{Left: rescaled proper energy density $\bar\rho_\phi$ and pressure $\bar{P}_\phi$ of the massless scalar field. 
    Right: the radial derivative of the rescaled pressure $\partial_{\bar{r}} \bar{P}_\phi$. In both panels, a solid line indicates that the value of the quantity is positive, while a dashed line indicates that it is negative. The lines are plotted using the 2-2-hole solution  with $r_H/\lambda_2=100$. The two vertical dotted lines denote $r=r_F$ and $r_H$ from left to right.}
    \label{fig:scalar pressure}
\end{figure}

Finally, let us discuss the circumvention of the no-scalar-hair theorems for 2-2-holes in the test field limit, following the improved proof of Bekenstein in Ref.~\cite{Bekenstein:1995un}. The proof is based on a careful analysis of the pressure of the scalar field (i.e. $T_r^r$ in Ref.~\cite{Bekenstein:1995un}) and its radial derivative by utilizing the conservation law of scalar field (i.e. $ \nabla _{\mu}{T^{\mu r}_\phi}=0$) at $r\geq r_H$. For comparison, we display in Fig.~\ref{fig:scalar pressure} the rescaled proper energy density, the pressure and its radial derivative.

We can first focus on the exterior, i.e. $r\gtrsim r_H$, where 2-2-holes most closely resemble black holes. The main distinction from Bekenstein's argument is that $P_\phi$ is positive at $r$ sufficiently near the horizon for 2-2-holes, while it is negative for black holes. This difference can be attributed to a boundary term $\propto \sqrt{B}P_\phi$, which is zero at the horizon for black holes assuming a finite $P_\phi$, and becomes significant as $P_\phi$ goes large when approaching the would-be horizon of 2-2-holes. Therefore, it is possible to have $P_\phi>0$ and $\partial_r P_\phi<0$ for all $r\gtrsim r_H$, without resulting in a contradiction with the conservation law as in the case of black holes.

However, as we delve into the interior of the 2-2-hole, $P_\phi$ does become negative and $\partial_r P_\phi$ turns positive at $r< r_F$. At first glance, the latter seems to contradict the conservation law of the scalar field. However, it is important to remember that the scalar field interacts with the matter source in this regime, and thus the conservation law applies only to their combination. As shown in Fig.~\ref{fig:backreaction}, the stress tensor of the scalar field is negligibly small compared to that of the matter source for $\gamma^2\lesssim \mathcal{O}(1)$. Therefore, the stress tensor of the matter source satisfies the conservation law at the leading order, while the scalar field and small perturbation of the matter source together obey the law at the next-leading order. We have confirmed that $ \nabla _{\mu}{T^{\mu r}_\phi}$ is indeed nonzero at $r\lesssim r_F$ from the numerical solutions,\footnote{In relation to the scalar field EOM in Eq.~(\ref{eq:phiEOM0}), the non-zero value of $ \nabla _{\mu}{T^{\mu r}_\phi}$ arises from the additional $r$ dependence introduced by $T(r)$ or $k_F(r)$ in either $V_T(\phi)$ or $V_\rho(\phi)$.} and thus the derivation in Ref.~\cite{Bekenstein:1995un} based on $ \nabla _{\mu}{T^{\mu r}_\phi}=0$ does not apply. This provides a concrete demonstration of how the no-scalar-hair theorems can be avoided by UCOs with a black hole-like exterior but a highly curved and matter-enriched interior.


\clearpage
\newpage

\bibliographystyle{unsrt}
\bibliography{ref_22holescalar}

\end{document}